\documentclass[]{aa}

\usepackage{graphicx}% Include figure files
\usepackage{dcolumn}% Align table columns on decimal point
\usepackage{bm}% bold math
\usepackage{placeins}
\usepackage{txfonts}
\usepackage{natbib}
\usepackage{amsmath}    % need for subequations
\usepackage{graphicx}   % need for figures
\usepackage{verbatim}   % useful for program listings
\usepackage{color}      % use if color is used in text
\usepackage{subfigure}  % use for side-by-side figures
\usepackage{gensymb} %for degree symbols and others
\usepackage{wasysym} %for astro symbols
\usepackage{hyperref}   % use for hypertext links, including those to external documents and URLs
\bibpunct{(}{)}{;}{a}{}{,} % to follow the A&A style
\usepackage{graphicx}
\usepackage{nth}
\newcommand{\kms}{\mbox{${\rm km\,s}^{-1}$}}

\newcommand{\Mjup}{\mbox{${M}_\mathrm{J}$}}

\newcommand{\Rjup}{\mbox{${R}_\mathrm{J}$}}

\begin{document}

\title{Detection of a high-velocity sodium feature on the ultra-hot Jupiter WASP-121~b\thanks{Based on Guaranteed Time Observations collected at the European Southern Observatory under ESO programme 1102.C-0744 by the ESPRESSO Consortium.
}}
    
\author{J.~V.~Seidel\inst{1} 
\and F.~Borsa\inst{2} %0000-0003-4830-0590  francesco.borsa@inaf.it
\and L.~Pino\inst{3} 
\and D.~Ehrenreich\inst{4}
\and M.~Stangret\inst{5} %comments to come
\and M.~R.~Zapatero~Osorio\inst{6}
\and E.~Palle\inst{5}
%from here alphabetical
\and Y.~Alibert\inst{7}%0000-0002-4644-8818  yann.alibert@space.unibe.ch comments to come
\and R.~Allart\inst{4,8}  %0000-0002-1199-9759 
\and V.~Bourrier\inst{4}
\and P.~Di~Marcantonio\inst{9}%0000-0003-3168-2289 paolo.dimarcantonio@inaf.it
\and P.~Figueira\inst{4,10} %0000-0001-8504-283X
\and J.~I.~Gonz\'{a}lez~Hern\'{a}ndez\inst{5} %jonay.gonzalezhernandez@iac.es  0000-0002-0264-7356 comments to come
\and J.~Lillo-Box\inst{6} %jlillo@cab.inta-csic.es 0000-0003-3742-1987
\and C.~Lovis\inst{4}
\and C.~J.~A.~P.~Martins\inst{10,11}%0000-0002-4886-9261 carlos.martins@astro.up.pt
\and A.~Mehner \inst{1} %0000-0002-9564-3302 amehner@eso.org
\and P.~Molaro \inst{9, 13} %0000-0002-0571-4163 paolo.molaro@inaf.it
\and N.~J.~Nunes\inst{14} %0000-0002-3837-6914 njnunes@fc.ul.pt
\and F.~Pepe\inst{4}
\and N.~C.~Santos\inst{10,11}
\and A.~Sozzetti\inst{15}  %0000-0002-7504-365X
}
%ADDTHIS check the portuguese affiliations, total mess
\institute{European Southern Observatory, Alonso de C\'ordova 3107, Vitacura, Regi\'on Metropolitana, Chile
\and INAF -- Osservatorio Astronomico di Brera, Via E. Bianchi 46, 23807 Merate (LC), Italy
\and INAF - Osservatorio Astrofisico di Arcetri, Largo E. Fermi 5, I-50125, Florence, Italy
\and Observatoire astronomique de l'Universit\'e de Gen\`eve, Chemin Pegasi 51b, 1290 Versoix, Switzerland
\and Instituto de Astrof\'isica de Canarias, Via Lactea sn, 38200, La Laguna, Tenerife, Spain
\and Centro de Astrobiolog\'ia (CSIC-INTA), Ctra. de Ajalvir km 4, 28850, Torrej\'{o}n de Ardoz, Madrid, Spain
\and Physikalisches Institut \& NCCR PlanetS, Universit\"{a}t Bern, CH-3012 Bern, Switzerland
\and Department of Physics, and Trottier Institute for Research on Exoplanets, Universit\'e de Montr\'eal, Montr\'eal, H3T 1J4, Canada
\and  INAF- Osservatorio Astronomico di Trieste, via Tiepolo 11, I-34143 Trieste, Italy
\and Instituto de Astrof\'{\i}sica e Ci\^encias do Espa\c co, CAUP, Universidade do Porto, Rua das Estrelas, 4150-762 Porto, Portugal
\and Centro de Astrof\'{\i}sica da Universidade do Porto, Rua das Estrelas, 
4150-762 Porto, Portugal
\and Departamento de F\'isica e Astronomia, Faculdade de Ci\^encias, Universidade do Porto, Rua do Campo Alegre, 4169-007 Porto, Portugal
\and Institute for Fundamental Physics (IFPU), Via Beirut 2, 34151 Grignano TS, Italy
\and Instituto de Astrof\'{i}sica e Ci\^{e}ncias do Espa\c{c}o, Faculdade de Ci\^{e}ncias da Universidade de Lisboa, Campo Grande, PT1749-016 Lisboa, Portugal
\and INAF - Osservatorio Astrofisico di Torino, Strada Osservatorio, 20 I-10025 Pino Torinese (TO), Italy
}

\date{Received date/ Accepted date}

\abstract{\textit{Context.} Ultra-hot Jupiters, with their high equilibrium temperatures and resolved spectral lines, have emerged as a perfect testbed for new analysis techniques in the study of exoplanet atmospheres. In particular, the resolved sodium doublet as a resonant line has proven a powerful indicator to probe the atmospheric structure over a wide pressure range.\\ 
\textit{Aims.} We explore an atmospheric origin of the observed blueshifted feature next to the sodium doublet of the ultra-hot Jupiter WASP-121~b, using a partial transit obtained with the 4-UT mode of ESPRESSO. We study its atmospheric dynamics visible across the terminator by splitting the data into mid-transit and egress. \\ 
\textit{Methods.} We explore the impact of the Rossiter-McLaughlin effect on the line shape of the sodium doublet. The partial transit is separated into one dataset centred around mid-transit and one dataset comprising the second part of the transit and egress. Lastly, the atmospheric retrieval code MERC is applied to both datasets in order to study the imprint of atmospheric dynamics on the line shape of the sodium doublet. \\ 
\textit{Results.} We determine that the blueshifted high-velocity absorption component is generated only during the egress part of the transit when a larger fraction of the day side of the planet is visible. For the egress data, MERC retrieves the blueshifted high-velocity absorption component as an equatorial day-to-night side wind across the evening limb, with no zonal winds visible on the morning terminator with weak evidence compared to a model with only vertical winds. For the mid-transit data, the observed line broadening is attributed to a vertical, radial wind.\\ 
\textit{Conclusions.} We attribute the equatorial day-to-night side wind over the evening terminator to a localised jet and restrain its existence between the substellar point and up to $10^\circ$ to the terminator in longitude, an opening angle of the jet of at most $60^\circ$ in latitude, and a lower boundary in altitude between [1.08, 1.15] $R_p$. As a hypothesis, we propose that the jet is produced by the excitation of standing planetary scale Rossby waves by stellar irradiation and subsequently broken by Kelvin-Helmholtz instabilities. Due to the partial nature of the transit, we cannot make any statements on whether the jet is truly super-rotational and one-sided or part of a symmetric day-to-night side atmospheric wind from the hotspot.
}


\maketitle

%----------------------------------------------------------------------------------------
%       ARTICLE CONTENTS
%----------------------------------------------------------------------------------------
\section{Introduction}

The atmospheric dynamics of our gas giant neighbours Jupiter and Saturn and their banded structure have been studied since the dawn of telescopes to better understand Earth's uniqueness \citep{Kaspi2020}. With the clear advantage of fly-bys of human-made spacecraft \citep{Smith1979} and in-situ probes such as the latest mission to Jupiter with Juno \citep{Kaspi2020}, the deep structure of Jupiter's atmosphere is remarkably well understood. However, to fully appreciate the atmospheric dynamics of gas giants, the observations of our closest planetary neighbours have to be seen in the wider context of population studies of exoplanets. \\

\noindent While low resolution studies from space have paved the way with indirect wind measurements \citep{Knutson2007,Cowan2007}, ground-based observations quickly caught up: the 2~\kms{} blueshift observed in high-resolution for the CO band of HD209458~b can be credited as the first ground-based attempt to directly measure atmospheric winds in an exoplanet atmosphere \citep{Snellen2010}. They compared the observed shift with the systemic velocity of the system and concluded that the blueshift is most likely induced by a day-to-night side wind dominating the pressure levels probed by the CO band (based on theoretical work by \citet{Knutson2008}). However, this first attempt relied on integrating over the entire dataset to compensate for a low signal-to-noise ratio (S/N) leaving many degenerate solutions to the observed blueshift, most notably a possible explanation of the shift from the uncertainty of the planetary orbit \citep{Montalto2011}.  

Recently, the Balmer lines have been used to probe winds beyond the thermosphere \citep[see e.g. ][for the study of Parker winds in KELT-9~b]{Wyttenbach2020}. In \cite{Cauley2020}, the offset of the line center of the Balmer lines was mapped throughout the transit, allowing for a first-order tracing of the wind speed as a function of time. They attribute the observed shift to large zonal winds and interpret this as evidence for day-to-night side winds, as they rule out a jet stream with elevated wind speeds. Most of these studies are conducted on hot or ultra-hot Jupiters, where the degeneracy between planetary rotation and atmospheric circulation can be broken as they are tidally locked \citep{Louden2015,Brogi2016}.

One of the most powerful observational probes of atmospheric dynamics as a function of time in the lower atmosphere is the tracking of the atmospheric Doppler shift via the cross-correlation technique \citep{Snellen2010} as first presented in \citet{Borsa2019,Ehrenreich2020}. Here the offset from the theoretical position in velocity space is traced throughout the transit for the forest of spectral iron lines which gives us additional spatial information on different visible parts of the atmosphere \citep[see also][]{Kesseli2021,Wardenier2022}. To probe a wider range of atmospheric layers in altitude, the sodium doublet, as a resonant line, has emerged as a readily available probe \citep{Wyttenbach2015}, where the Doppler shift induced change in line shape can be attributed to atmospheric dynamics, e.g. via MERC \citep[a Multi-nested ETA Retrieval Code, ][]{Seidel2020,Seidel2021, Mounzer2022}. While able to distinguish between different winds patterns based on retrievals for the first time, these works also relied on the time integration over the entire transit and could not distinguish any evolution of wind patterns based on different viewing geometries of the exoplanet atmosphere. Nonetheless, the sodium retrieval technique spanning orders of magnitude in altitude and the time-resolved CCF tracing of the lower atmosphere independently deduced a day-to-night side wind of approximately $5~\kms{}$ in the lower atmosphere of the ultra-hot Jupiter WASP-76~b \citep{Ehrenreich2020,Seidel2021} with an outwards bound wind linking to higher atmospheric layers \citep{Seidel2021}. 

In this work, we study the atmospheric dynamics of WASP-121~b with MERC as a first exploration into time-resolving narrow-band transmission spectroscopy. WASP-121~b is a highly irradiated exoplanet orbiting a relatively bright F6V-type star (V=10.4 mag, $T_{\mathrm{eq}} = 2358 \pm 52K$) with a bloated atmosphere ($R = 1.753\pm 0.036\Rjup, M = 1.157\pm0.070\Mjup$) \citep{Delrez2016,Bourrier2020}. Given its closeness to its host star ($0.02544 \pm 0.0005$ AU) and the resulting high equilibrium temperature ($2358\pm52~K$), it is classified as an ultra-hot Jupiter \citep{Evans2017}, a class of exoplanets most notably characterised by molecular dissociation and the resulting atomic spectral lines \citep{Arcangeli2018, Kitzmann2018, Lothringer2018, Parmentier2018}. From H$_2$O observations with the WFC3 and STIS instruments on the \textit{Hubble Space Telescope} (HST) \citet{Evans2017} inferred a temperature inversion in its day-side atmosphere. The temperature profile of both the night and day-side of this tidally-locked exoplanet was then constrained in \citet{MikalEvans2022} and \citet{Daylan2021}. They observe a difference in temperature between the day and night side of the planet of $1000~K$ with approximately $1500~K$ on the cooler night side and $2500~K$ on the hotter day side which then increases to over $4000~K$ above $10^{-3}$~bar. However, there is no significant phase shift between the brightest and sub-stellar points which was confirmed from TESS optical photometry in \citet{Bourrier2020}. This indicates inefficient heat transportation across the terminator and two distinct atmospheric regimes in the day and night side realm. In \citet{Maguire2022}, high-resolution ESPRESSO\footnote{Echelle SPectrograph for Rocky Exoplanets and Stable Spectroscopic Observations} data, including the here studied 4-UT transit, is used to retrieve relative abundances and temperature pressure profiles. They find $3450\pm160~K$ for pressures above $10^{-2}$~bar compatible with the previously mentioned temperature inversion, and also in line with UVES observations in \citet{Gibson2020,Gibson2022}.

\noindent A wide range of elements were identified in the transmission spectrum of WASP-121~b using ground based data  \citep{Bourrier2020,Bourrier2020b,Cabot2020,Hoeijmakers2020,Gibson2020,Merritt2021,Borsa2021}, for our work most importantly the resolved sodium lines. Thanks to advances in high-resolution cross-correlation techniques (CCF), the non-detection of Ti and TiO \citep{Hoeijmakers2020,Merritt2021,Wilson2021} was attributed to a depletion of Ti at the terminators \citep{Gibson2020}. Recently, \citet{Hoeijmakers2022} interpreted these trends as cold-trapping of Ti on the night side of the atmosphere, which implies a wider variety in ultra-hot Jupiter atmospheres based on slight differences in temperature or dynamical structure. Applying the cross-correlation technique to iron on a HARPS dataset of WASP-121~b showed a blue offset of $5.2\pm0.5~\kms{}$ which could trace a day-to-night side wind deep in the atmosphere and confirmed its tidally-locked nature \citep{Bourrier2020b}. These time-integrated results are  in agreement with \citet{Borsa2021} where the CCF iron signature is traced in time. They find a blueshift of $-2.80\pm0.28 \kms{}$ for mid-transit that then increases to $-7.66\pm0.16 \kms{}$ for the second half of the transit indicative of either a super-rotational wind or a day-to-night side wind transporting material towards the observer over both limbs in the deep atmosphere.  \\

We utilise the superior S/N of ESPRESSO's 4-UT mode to study a distortion in the line shape of the narrow-band transmission spectrum of the sodium doublet resolved in time at first order. We show that the resulting high-velocity shifted feature in the blue arm of the sodium doublet is not an artefact but of atmospheric origin and attribute it to a localised jet-stream that comes into view during the second part of the transit. In Section \ref{sec:dataset}, we discuss the reanalysis of the ESPRESSO 4-UT data compared with the first publication in \citet{Borsa2021} and the separation of the transit in two datasets. Section \ref{sec:MERC} highlights the main functionalities of MERC in addition to the new implementation of localised jets, Section \ref{sec:results} discusses the retrieval results on the two distinct datasets centred at mid-transit and egress. Lastly, we put our work in context of current literature in Section \ref{sec:diss}.

\section{ESPRESSO dataset and reanalysis}
\label{sec:dataset}

We analyse the line shape of the sodium doublet of WASP-121~b observed with ESPRESSO. ESPRESSO, a fibre-fed, ultra high-resolution, stabilized echelle spectrograph, has the unique capability to receive light from each Unit Telescope (UT) of the Very Large Telescope (VLT) at ESO's Paranal observatory in Chile \citep{Pepe2021}. Each UT has an 8.2m mirror, combining light from all four UTs results in the equivalent of a 16m-class telescope observation. This unique 4-UT mode (mid-resolution, binning 4x2: MR42) has a resolution of R$\approx70,000$ and boasts a significant increase in S/N when compared to the higher resolution 1-UT mode for the same exposure time (see \citet{Borsa2021} for a comparison between 4-UT and 1-UT mode observations). 
\noindent In this work, we re-analyse the partial 4-UT transit of WASP-121~b covering mid-transit and egress which was already presented in \citet{Borsa2021}, starting at phase $-0.017$ before mid-transit until egress at phase $0.045$. The partial transit was obtained as part of the commissioning of the 4-UT mode on the 30-Nov-2018 and processed with the ESPRESSO pipeline v2.8.0. For an in-depth discussion of the observing conditions and the mode, see \citet{Borsa2021}. The average S/N at 550nm was 180 with an exposure time of 300s. One of the most important parameters in this study is the mean systemic velocity. We use the value as derived in \citet{Borsa2021}, which is in agreement with the value reported in the detection paper of this system \citep{Delrez2016}.

\noindent Our focus lies on the line shape of the sodium doublet to study the atmospheric dynamics of WASP-121~b with an emphasis on the secondary high-velocity absorption components seen on the blueshifted side of the sodium lines at approximately $5889.2 \AA$ and $5895.2 \AA$.

\subsection{Time resolving the transit}

\begin{figure}[htb]
\resizebox{\columnwidth}{!}{\includegraphics[trim=1.0cm 2.0cm 0.0cm 1.0cm]{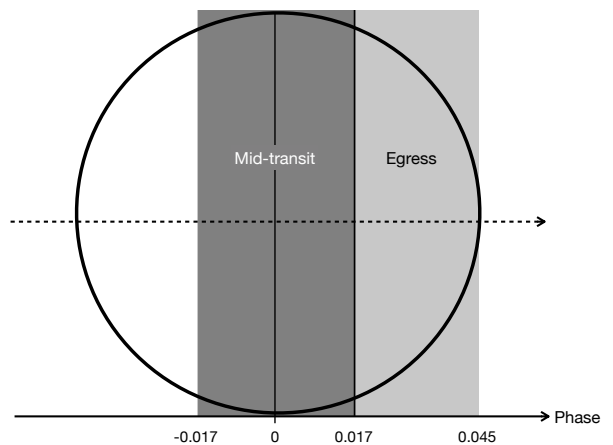}} 
	\caption{Graphic of the data taken and the split in two datasets for the 4-UT ESPRESSO transit. Time is indicated as phase on the x-axis, with the planetary orbit marked as a dashed arrow indicating the direction of movement in time across the stellar disk. The gray areas indicate when exposures where obtained during transit and how the data was split into one set centred around mid-transit (0.0, darker gray, mid-transit dataset) and the other comprising the rest of the exposures towards egress (from here called the egress dataset).}
	\label{fig:transit_overview}
\end{figure}

\noindent The sodium doublet in WASP-121~b is shallower than in comparable ultra-hot Jupiters \citep[e.g. in WASP-76~b ][]{Seidel2019,Tabernero2020} and resolving the line shape is challenging. For this reason, no definitive conclusion about the line ratio or broadening could be drawn from a dataset of three HARPS nights in \citet{Hoeijmakers2020} or from the 1-UT data in \citet{Borsa2021}. The 4-UT data, covering the transit only partially, allow for a superior S/N and resolved spectral lines, despite the moderately reduced resolution of the 4-UT mode compared to ESPRESSO used in 1-UT mode. During transit, 19 exposures were taken, starting at orbital phase $-0.017$. For symmetry, this means a set of 10 exposures is centred around mid-transit from orbital phases $-0.017$ to $0.017$, while a set of 9 exposures is left for the egress part of the transit. An overview of the coverage of the data taken as well as the split into two datasets is shown in Figure \ref{fig:transit_overview}. A finer split of the data in smaller subsets is not feasible to achieve convergence of MERC while also maintaining symmetry around mid-transit and maintaining subsets of similar number of exposures. The egress dataset contains orbital phases $0.017$ to $0.048$ (end of transit). To build the transmission spectrum we follow the steps outlined in \citet{Seidel2022}. To separate the stellar spectral lines from the planetary spectrum, we divide each dataset by the master-out spectrum consisting of out-of-transit exposures only. For the master-out spectrum, ten out-of-transit exposures were available, of which we use seven. The first three exposures after egress that still contain the planet's Hill sphere are not used in our analysis to avoid contamination of a possible atmospheric tail, even if no such tail was detected in the exposures in question due to insufficient S/N.

\noindent Inhomogenities in the Coudé Train optics leading to ESPRESSO induce interference patterns in each spectrum \citep{Sedaghati2021}. As a consequence, the transmission spectrum contains sinusoidal noise (wiggles) in the continuum \citep{Allart2020,Tabernero2020}. A sinusoidal fit excluding the wavelength region on the sodium lines is performed to correct for this feature following \citet{Seidel2022}. For more information on this interference pattern, see Section 5.3 of the ESPRESSO user manual\footnote{\url{https://www.eso.org/sci/facilities/paranal/instruments/espresso/ESPRESSO_User_Manual_P109_v2.pdf}}.

\begin{figure}[htb]
\resizebox{\columnwidth}{!}{\includegraphics[trim=0.0cm 1.0cm 0.0cm 1.0cm]{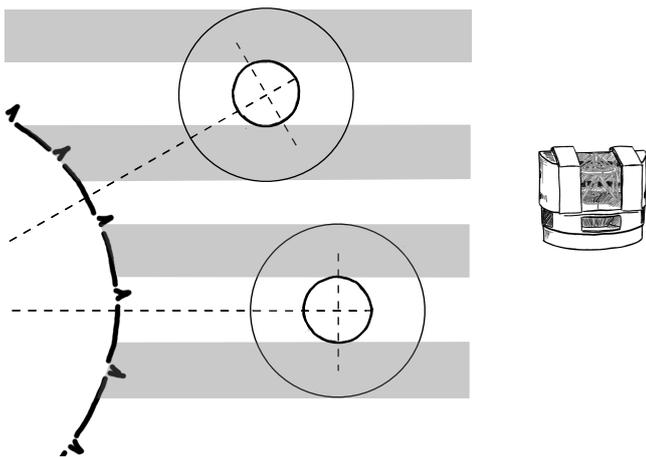}} 
	\caption{Graphic of atmospheric parts probed during the separated parts of mid-transit and egress data in a polar planetary view. The host star is indicated on the left, with the observer as a small VLT UT on the right. WASP-121~b is shown in the middle with its obscure planet disk as the smaller circle and the atmosphere extension as the larger circle. The grey areas indicate the light from the host star crossing the exoplanet atmosphere over each limb in the line of sight. The depiction of WASP-121~b in the bottom is shown roughly at mid-transit, where the terminator (short, dashed line) is orthogonal to the line of sight. In the top it is shown at a moment in time during the egress part of the dataset, when the planet has rotated and shows a different opening angle than during mid-transit, with different parts of the atmosphere probed across both limbs. The opening angle is indicated by the longer, dashed lines starting at the centre of the host star. The graphic is not to scale.}
	\label{fig:sketch_polar}
\end{figure}

\noindent During its transit, a tidally locked planet will have slightly different parts of its atmosphere illuminated by the host star across both limbs. The terminator, separating the day- from the night-side is only perfectly aligned orthogonally with the line of sight at mid-transit, while each exposure taken further away from mid-transit will show an inclination. This difference in viewing angle of the atmosphere is described by the opening angle of the atmosphere measured from mid-transit \citep[see][]{Wardenier2022} and shown in the sketch in Figure \ref{fig:sketch_polar} where the polar view of the system is shown with the view of one moment in time during egress on the top and one moment at mid-transit in the bottom. 

\begin{figure*}[htb]
\resizebox{\textwidth}{!}{\includegraphics[trim=0.0cm 6.0cm 0.0cm 7.0cm]{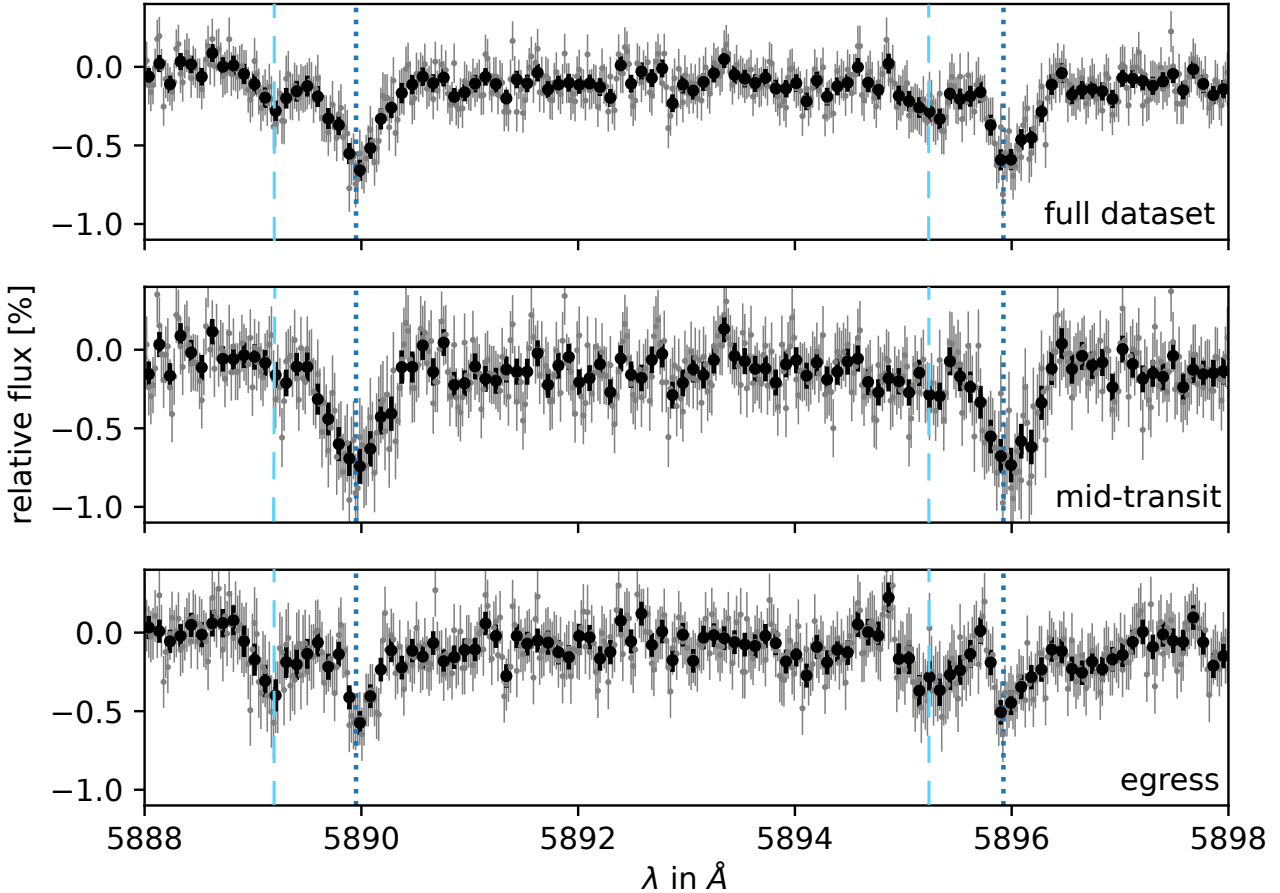}} 
	\caption{The sodium doublet transmission spectra for the partial 4-UT transit normalised to 0. The top panel shows the full dataset combined, in agreement down to the noise level with \cite{Borsa2021}, the central panel the mid-transit centred dataset, and the bottom panel the egress centred dataset. The line centre of the sodium doublet lines are indicated as dotted, blue, vertical lines. The egress data shows a distinct high-velocity absorption feature on the blue side of each sodium doublet line, which is the feature under investigation in this work. The position of this feature is highlighted in each of the three panels by the light-blue, dashed line.}
	\label{fig:data_overview}
\end{figure*}

From mid-transit, the opening angle at the start of the egress dataset is $\sim 10^\circ$, at the end of the egress dataset it is $\sim 30^\circ$, with the mean opening at $\sim 20^\circ$. A more thorough treatment of the opening angles of hot Jupiters during transit can be found in \citet{Wardenier2022}. Our re-analysis of the dataset when combining all spectra to one transmission spectrum for the transit is compatible to the noise level with the sodium doublet presented in \citet{Borsa2021} and is shown in Figure \ref{fig:data_overview}, top panel. The two split datasets in the wavelength range of the sodium doublet in the planetary rest frame are shown in Figure \ref{fig:data_overview} in the central panel for the mid-transit centred dataset and in the bottom panel for the egress centred dataset. The mid-transit sodium doublet exhibits a pronounced broadening, while the egress dataset sodium signal is split in a component at line centre (marked with a vertical, blue, dotted line) and a high-velocity blueshifted absorption feature (position marked with a light-blue, dashed line). In the following we will investigate the presence of this feature and possible atmospheric and contamination origins.

\subsection{Doppler smearing}

\noindent In the study of atmospheric dynamics from the line shape, one of the most treacherous pitfalls is the impact of the exposure time. The planetary movement itself during the exposure introduces smearing of the spectrum and thus produces an artificial broadening of the profile \citep{RiddenHarper2016, Wyttenbach2020, Cauley2020}. The longer the exposure time, the more pronounced this effect becomes. However, the exposure time also has to be long enough to provide a S/N high enough for the spectrum to be photon noise dominated and not readout noise dominated in the deepest stellar spectral features \citep[for more information see][]{BorsaZannoni2018, Seidel2020b}. Ideally, this trade-off is taken into account when the observations are prepared following Boldt-Christmas et al. (in prep). However, when the study of atmospheric dynamics from the line shape is a byproduct of the initial science case, the impact of the smearing has to be estimated and propagated through the analysis if non-negligible, e.g. by altering the data via convolution or by adjusting the Voight profiles that generate the model spectrum. The exposure time for the 4-UT mode partial transit was 300s. The readout between exposures for MR42 mode is 42~s, leading to a gap in coverage but no additional smearing. For WASP-121~b, this leads to a maximum smear of $1.1\, \kms$. In consequence, the smear in one exposure amounts to one resolution element in 4-UT mode. This has no impact on the conclusions, but the additional uncertainty on the broadening is accounted for.

%Doppler smearing: dRV_one exposure = DeltaRV_orb_transit/(T14*24.*3600.)*exposure_time.

\subsection{Impact of the Rossiter-McLaughlin effect}

\begin{figure}[htb]
\resizebox{\columnwidth}{!}{\includegraphics[trim=0.0cm 0.0cm 0.0cm 0.0cm]{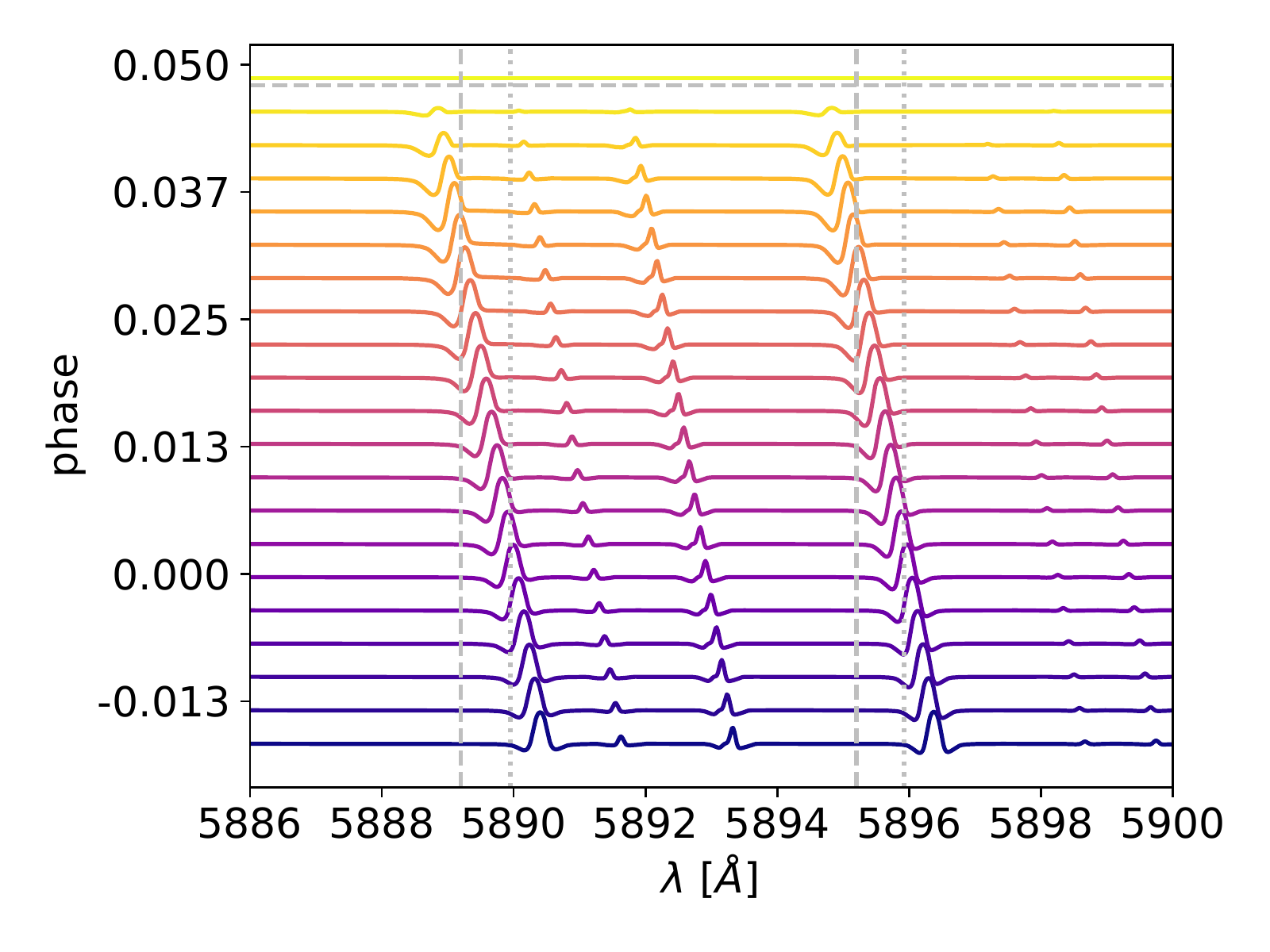}} 
	\caption{The RM+CLV model from \citet{Borsa2021} as a function of phase in the planetary rest frame. The start of the transit is at the bottom, time moves vertically and is emphasized by the colour gradient. The end of the transit is indicated with a dashed, grey horizontal line. The wavelength position of the sodium doublet is indicated as dotted, grey vertical lines, the centre of the high-velocity absorption components are indicated as dashed, grey vertical lines.}
	\label{fig:RM_phase}
\end{figure}

\noindent In-transit spectra are affected by the Rossiter-McLaughlin effect \citep[RM: ][]{Rossiter1924,Mclaughlin1924,Cegla2016}, a Doppler shift induced by the stellar rotation, and by the stellar center-to-limb variations (CLV), which describe the non-uniform surface brightness of the stellar disk. For consistency, we apply the exact RM+CLV model from \citet[][see details therein]{Borsa2021} which is based on the methodology proposed in \citet{Yan2017} and shown in Figure \ref{fig:RM_phase} for each in-transit spectra. In the following, we assess if the origin of the blueshifted high-velocity absorption components could stem from the RM+CLV effect or its over-correction. The amplitude of the model decreases for higher phases (see stacked models in phases in Figure \ref{fig:RM_phase}), indicating that the RM+CLV correction has a lesser impact on the egress dataset than on the mid-transit dataset. Nonetheless, the RM+CLV correction impacts the wavelength region of the high-velocity absorption components as well as the wavelength region of the main sodium doublet and has to be studied carefully to attribute the seen features to atmospheric processes. 

\begin{figure}[htb]
\resizebox{\columnwidth}{!}{\includegraphics[trim=0.0cm 0.0cm 0.0cm 0.0cm]{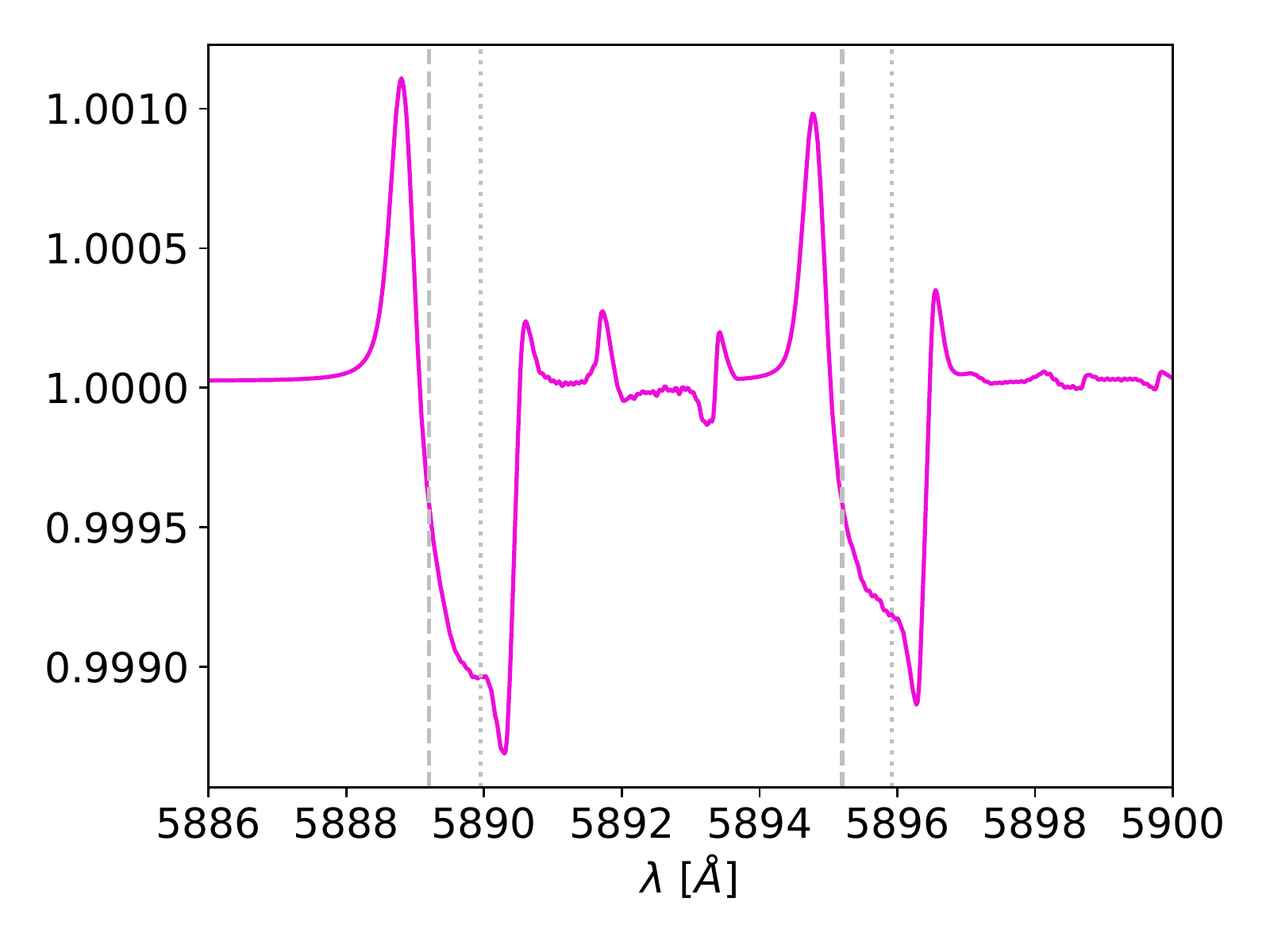}} 
	\caption{The propagation of the impact of the RM+CLV-models for each exposure from Figure \ref{fig:RM_phase} in the final transmission spectrum. The wavelength position of the sodium doublet is indicated as dotted, grey vertical lines, the centre of the high-velocity absorption components are indicated as dashed, grey vertical lines. The main impact of the RM+CLV correction is at the centre of the sodium doublet, only impacting the mid-centre data and on the blue side of the high-velocity absorption components.}
	\label{fig:RM_impact}
\end{figure}

\begin{figure*}[htb]
\resizebox{\textwidth}{!}{\includegraphics[trim=3.0cm 0.0cm 0.0cm 0.0cm]{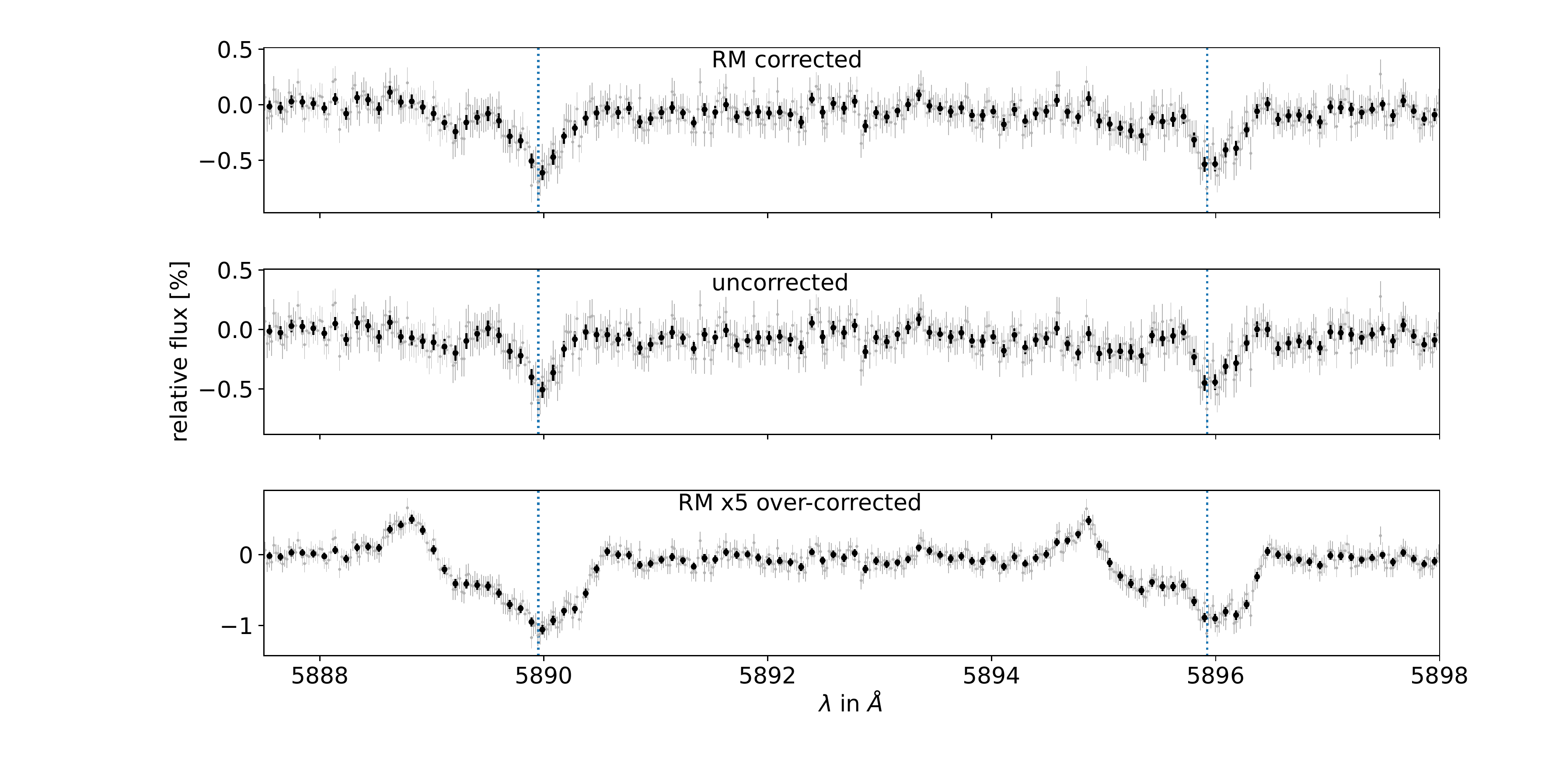}} 
	\caption{Comparison of the sodium doublet with the RM-effect corrected as in \citet{Borsa2021}, without corrections, and over-corrected by a factor of 5. The line position of the sodium doublet is indicated as blue, dotted lines. The corrected doublet is seen on top, the uncorrected doublet in the middle, and the over-correction on the bottom. The high-velocity absorption components cannot be explained by an incorrect RM-effect correction, but are of atmospheric origin as the feature is seen in the uncorrected, corrected and even smeared out in the over-corrected spectrum.}
	\label{fig:RM_impact_comparison}
\end{figure*}

\noindent To assess the strength of the combined RM-correction in the final transmission spectrum, we propagated the model correction through the transmission spectrum pipeline, which is shown in Figure \ref{fig:RM_impact}. The strongest change of the transmission spectrum due to the RM-correction is seen at the centre of the sodium doublet and on the blue side of the high-velocity absorption components (marked with a dotted grey line). The impact of that correction is visible in the transmission spectrum (compare the top and bottom panels of Figure \ref{fig:RM_impact_comparison}) where the RM-correction refines the wing of the high-velocity absorption components. 

\noindent Concluding from the RM-correction, the RM-effect distorts the blue wing of the high-velocity absorption components, which is refined again after the correction. We varied the amplitude of the RM-correction between no correction and five times the amplitude derived in \cite{Borsa2021} and were unable to erase the high-velocity absorption components. The only visible effect is a distortion of the blue wing of the high-velocity absorption components. This means that the feature on the blue side of both sodium lines is not an artefact of the RM-effect itself or its correction. An additional argument against an origin of the high-velocity absorption feature from the CLV effect stems from the visual inspection of other spectral lines that are also present in the stellar spectrum as deep spectral lines, e.g. the CaII H and K lines \citep[see Figure 10 in ][]{Borsa2021}. As shown for different ESPRESSO datasets, the CLV effect imprints on all stellar spectral lines in the same shift direction \citep{CasasayasBarris2021,CasasayasBarris2022}, albeit at different magnitudes due to its wavelength dependence. Neither the CaII H or K lines show any blue-shifted component for WASP-121~b, providing further evidence against a RM+CLV origin of the feature under investigation.

\subsection{Impact of stellar or telluric lines}

\noindent Two other factors that could potentially impact the line shape of the sodium doublet are an insufficient or over-correction of telluric lines, residuals of stellar lines, or telluric sodium emission. 
We corrected for telluric lines with {\tt molecfit} \citep{Smette2015, Kausch2015}, an ESO software which takes into account the site conditions at the time of observation. For more details on molecfit see \citet{Allart2017}. We applied the same convergence parameters as described in \citet{Seidel2019} and verified that all telluric lines were corrected down to the noise level in the master spectrum. No residual from the stellar line removal was observed and the S/N remained sufficiently high even at the line core of the stellar sodium doublet as to not warrant an exclusion of the affected spectra \citep[see][for a discussion of the impact of low S/N stellar residuals]{Borsa2019, Seidel2020b, Seidel2020c}. Additionally, in the transmission spectrum as shown e.g. in Figure \ref{fig:data_overview} no residual from the stellar Ni line at $5892.88~\AA$ is visible. 
The barycentric Earth's radial velocity ranged from $11.4$ to $11.1~\kms{}$ during the partial 4-UT transit, with the planetary velocity ranging from $-22.7$ to $86.1~\kms{}$. Therefore, residuals of stellar or telluric lines, as well as any impact from the interstellar medium, would be smeared significantly and most likely not show as a concentrated feature. As a consequence, we rule out these possible sources for the high-velocity absorption component.

Telluric sodium emission is normally monitored on sky via the fibre B of the instrument. However, for the 4-UT-mode transit, fibre B was used for simultaneous wavelength calibration on the Fabry-Perot and no sky observations are available. There are two possible sources of telluric sodium emission: contamination from the laser at UT4 or excitation of atmospheric sodium from meteors. In 4-UT-mode all four UT telescopes of the VLT are used for ESPRESSO and as a consequence the laser on UT4 is not active, ruling out any laser contamination. The observations took place on the 30th Nov 2018, and no meteor activity was recorded over Chile during that time period. The next meteor shower which could excite the atmosphere sufficiently, the Geminids, did not start before December 4th (peak Dec 14th). As a consequence we also rule out systematic telluric sodium emission as a possible source of the observed feature.

\section{Application of MERC}
\label{sec:MERC}

\noindent MERC (Multi-nested $\eta$ Retrieval Code) \citep{Seidel2020} compares a provided transmission spectrum with a 3D forward model via a nested-sampling retrieval algorithm. The forward model calculates the temperature-pressure profile in 1D and then adds 3D wind patterns and constant planetary rotation assuming tidal locking. The zonal wind patterns can either be constant over the entire atmosphere or scale with solid body rotation with stronger winds at the equator and no winds at the pole \citep{Seidel2021}. 

\subsection{Implementation of localised jets}
\label{sec:jet}

\begin{figure}[htb]
\resizebox{\columnwidth}{!}{\includegraphics[trim=1.0cm 0.0cm 0.0cm 0.0cm]{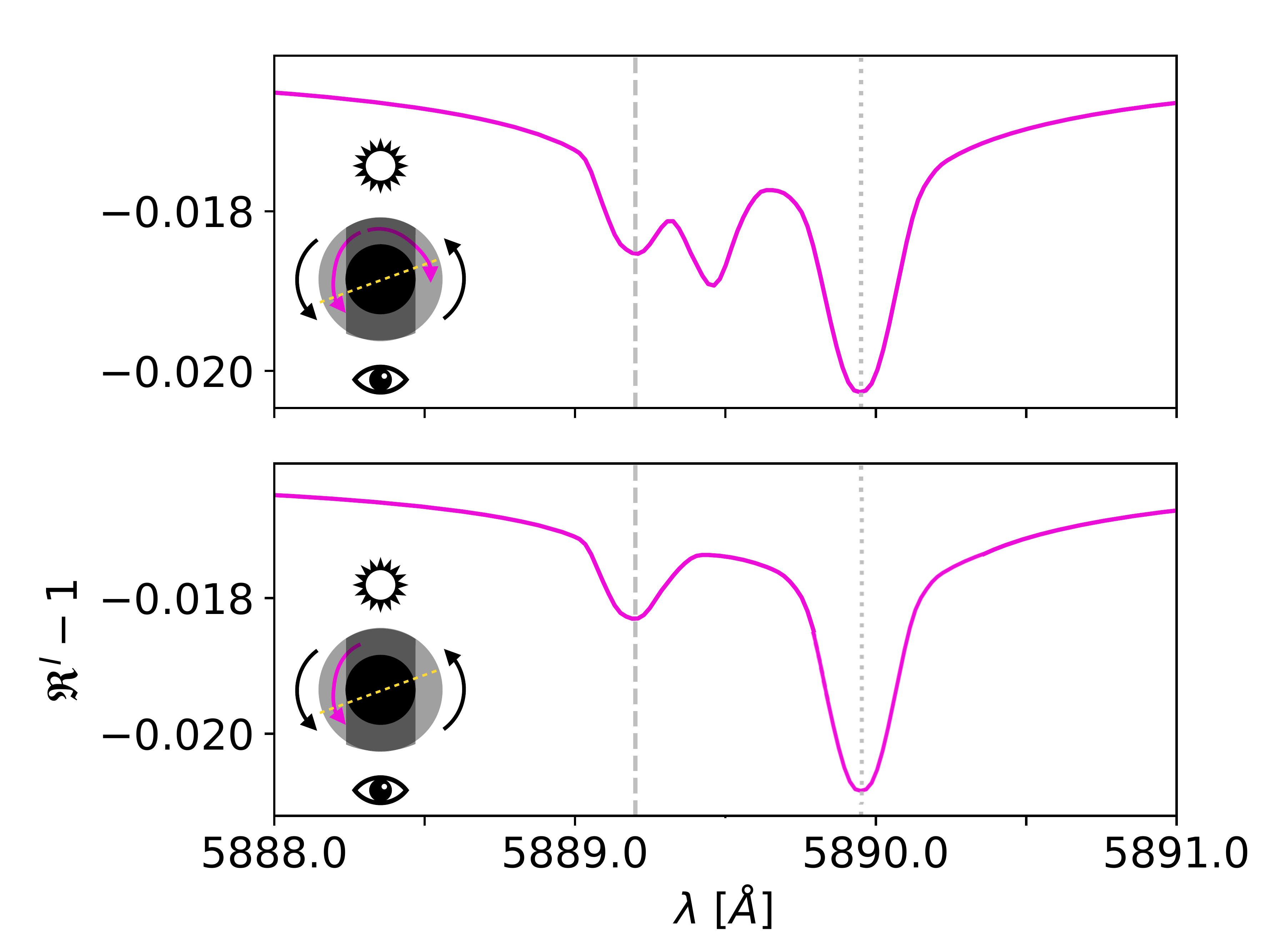}} 
	\caption{Two models generated with the wind model from MERC for the wavelength range of the sodium D$_2$ line. The position of the sodium line is marked with a dotted, gray vertical line. The position of the blueshifted high-velocity absorption components from the egress dataset are marked with a dashed, gray vertical line. The lower left corner of each panel shows a sketch of the atmospheric movements. The eye symbolises the observer, the sun the host star, the black arrows the planetary rotation over each terminator. The scarlet arrows show the direction and existence of atmospheric winds in the lower atmosphere and the yellow dashed line the terminator position to indicate the rotation of the planet with respect to the line of sight for the egress data.}
	\label{fig:jet_overview}
\end{figure}

\indent We study a possible atmospheric source for the blueshifted high-velocity absorption components at the sodium doublet. After ruling out a spurious origin of the feature, the high-velocity absorption component likely stems from the blueshift of parts of the sodium population in the atmosphere. 
The striking features of the high-velocity absorption component are both its distinguishable shape and its clear separation from the main sodium lines hinting at a physically distinct component. A global zonal wind would result in a broad range of velocities visible in the line of sight (LOS) which would in turn induce a gradual change of the Doppler shift when accounting for the entire atmosphere. As a consequence, the imprint on the transmission spectrum would show a wider range of shifts and a continuous impact on the line shape, like e.g. as seen for WASP-76~b with its asymmetric line shape encoding both broadening and a blueshifted absorption component from the lower atmosphere \citep{Seidel2021}. Instead of one continuous but deformed line, we see a bimodal distribution of a feature centred at the expected line centre, marking a symmetric distribution of Doppler shifts in the blue and red, and a separate feature at a high velocity range, with no connecting intermediate velocity range detected. This is indicative that the particle movement is likely not extending over the entire hemisphere where it would be strongest at the equator but gradually diminish towards the poles over a wide range of latitudes, but that the moving part of the atmosphere is rather localised, in a jet-like stream leaving us with an imprint of only high velocity Doppler shifts.
\noindent If the high-velocity absorption components were a consequence of a super-rotational jet that spans the entire atmosphere, it should have a redshifted equivalent from the evening terminator. Each side of the sodium line would have one high-velocity absorption component from the planetary winds aligned with the planetary rotation, creating a symmetric feature around each sodium line. If we consider the second option that can produce strong blue shifts, day-to-night side winds, we are left with the two cases shown in Figure \ref{fig:jet_overview}. Assuming the day-to-night side wind streams across both terminators equally, on the morning terminator against the planetary rotation and on the evening terminator with the direction of the planetary rotation, we would get two high-velocity absorption components (top panel). One at the position of the observed high-velocity absorption component of the egress data and one offset by the differential of the planetary rotation (two times the rotational velocity). If, due to the opening angle of the planet, only one side of the day-to-night side is visible across the terminator, namely the evening terminator, we have the situation of the bottom panel, where only one stream is visible, creating one blueshifted high-velocity absorption component. 

It is that scenario which was subsequently implemented in MERC as an additional wind model for the egress data. In the conclusions in Section \ref{sec:diss} we discuss all different degenerate scenarios that would generate the same imprint on the line shape as discussed above and make the case why a one-sided jet from the day to the night side remains the most likely scenario. To simulate a more locally contained jet, we implement the day-to-night side wind solely in the latitude range of the tropics and sub-tropics, mimicking observed jets in Solar System planets \citep{Tollefson2017,Kaspi2020}. In MERC, each atmospheric quadrant is divided in nine atmospheric slices for convergence \citep[see][for an in-depth explanation of the setup of the model atmosphere]{Seidel2021}. We subsequently implement day-to-night side winds only in the three slices surrounding the equator, leading to an opening angle of $60^\circ$ for the jet and keep its $\cos\theta$ dependence. Given that a true turning of the model atmosphere to have less day side morning terminator visible during egress is not feasible in the current setup and subsequently left for future work, we have switched off the lower atmospheric wind on the morning terminator, creating a model with one high-velocity absorption component for each sodium doublet line.

\subsection{Model comparison}

\begin{table*}
\caption{Overview of the different models' prior ranges.}
\label{table:priors}
\centering
\begin{tabular}{l c c c c c}
\hline
\hline
Model & T$_{\mathrm{iso}}$ [K]  &  NaX & v$_{\mathrm{srot}}$ [\kms] & v$_{\mathrm{dtn}}$ [\kms] & v$_{\mathrm{ver}}$ [\kms]     \\
\hline
isothermal  & [1500, 5000]    & [-5.0, -1.0]  & -  & -  &  - \\ 
$\mathrm{dtn}_{\cos\theta}$  &  [1500, 5000]   & [-5.0, -1.0]  & -  &  [0.1,50.0]  & -\\
$\mathrm{srot}_{\cos\theta}$  &  [1500, 5000]   & [-5.0, -1.0]  & [0.1,50.0]   &  - & - \\
ver  &   [1500, 5000]  & [-5.0, -1.0]  & -  &  - & [0.1,50.0] \\
$\mathrm{jet}_{\cos\theta}, \mathrm{ver}$ &   [1500, 5000]  & [-5.0, -1.0]  & -  &  [1.0,50.0] & [0.1,50.0] \\
\hline
\end{tabular}
\end{table*}
For each dataset, the algorithm calculates the Bayesian evidence $|\ln\mathcal{Z}|$ of the current model over the parameter space provided by the priors. The used priors in this work are shown in Table \ref{table:priors}. The best-fit parameters are the median of the marginalised posterior distributions. With this approach, different models fitted to the same dataset can be compared by the difference in their Bayesian evidence $|\ln\mathcal{B}_{01}|$. We can subsequently judge the significance of the selection of one model over another via the Jeffreys' scale (see Table \ref{table:Bayesianoverview} or \citet{Trotta2008} and \citet{Skilling2006}). The overview of the Jeffreys' scale, with a more in depth explanation of its implications and implementation details of MERC can be found in \cite{Seidel2020}.

\begin{table}[bh]
\caption{The 'Jeffreys' scale'. It is an empirical scale to compare two model fits ($M_0$ and $M_1$) to the same data via the Bayesian evidence.}
\label{table:Bayesianoverview}
\centering
\begin{tabular}{c c c l }
\hline
\hline
$|\ln\mathcal{B}_{01}|$   & Odds & Probability & Strength of evidence     \\
\hline
$<1.0$  &  $<3:1$   &  $<0.750$  &   Inconclusive \\
$1.0$  &  $\sim 3:1$   &  $0.750$  &  Weak evidence \\
$2.5$  &  $\sim 12:1$   &  $0.923$  &   Moderate evidence \\
$5.0$  &  $\sim 150:1$   &  $0.993$  &   Strong evidence \\
   \hline
\end{tabular}
\end{table}

\section{Retrieval results}
\label{sec:results}

We applied MERC to both datasets independently, mid-transit and egress as shown in the top and bottom panel of Figure \ref{fig:dtn_verbestfit} in its updated form with planetary rotation and solid body rotation as described in \citet{Seidel2021}. The model applied to the egress data only is the jet as described in Section \ref{sec:jet} in the lower atmosphere combined with vertical winds in the upper atmosphere with different altitudes explored for the layer change in this section. The vertical wind is included for the egress dataset based on the results from the mid-transit dataset which will be discussed later on and accounts for the line broadening of the central line component. A list of all the models fed into MERC is shown in the following:

\begin{itemize}
\item isothermal model with no atmospheric dynamics apart from planetary rotation
\item $\cos\theta$ dependent day-to-night side wind ($\mathrm{dtn}_{\cos\theta}$)
\item $\cos\theta$ dependent super-rotational wind ($\mathrm{srot}_{\cos\theta}$)
\item vertical wind (ver)
\item egress only: two layer approach with a jet towards the observer on the evening terminator in the lower atmosphere combined with a vertical wind in the upper atmosphere. ($\mathrm{dtn}_{\cos\theta}, \mathrm{ver}$)
\end{itemize}

One of the most powerful tools in Bayesian retrieval is the application of priors. Priors quantify the prior knowledge of the studied problem in boundary conditions. The priors for each of the tested models are listed in Table \ref{table:priors}. The prior on the degenerate continuum parameter NaX was left to span various orders of magnitude to allow for a wide range of pressures, given that the pressure and sodium abundance have an important impact on the broadening. This parameter, which is explained in detail in \citet{Seidel2020}, parametrises the degeneracy between the pressure and the abundance that arises from the loss of the planet's continuum. This loss of absolute flux value is a result of the processing of high-resolution ground-based data and is explored in more depth in \citet{Birkby2018}. The prior of the temperature range is set taking into account the results from \citet{MikalEvans2022} spanning from the temperature of the cooler night side to the highest temperature of the temperature inversion.

\subsection{Application to mid-transit data}

\begin{table}
\caption{Comparison of the different models for the mid-transit data.}
\label{table:comparison_mid_transit}
\centering
\begin{tabular}{l c c l}
\hline
\hline
Model & $|\ln\mathcal{Z}|$   &  $|\ln\mathcal{B}_{01}|$ & Strength of evidence  \tablefootmark{a}  \\
\hline
$\mathrm{dtn}_{\cos\theta}$  &  $1112.17\pm0.11$   & -  & - \\
$\mathrm{srot}_{\cos\theta}$  &  $1113.87\pm0.10$   & $1.70$  & Weak evidence\\
isothermal  &  $1115.37\pm0.08$   & $3.20$  & Moderate evidence \\
\textbf{ver}  &  $1118.13\pm0.08$   & $5.96$  & Strong evidence \\
\end{tabular}
\tablefoot{\tablefoottext{a}{ The base model to calculate $|\ln\mathcal{B}_{01}|$ is the isothermal temperature model with $\mathrm{dtn}_{\cos\theta}$ winds. The comparison is based on the Jeffreys' scale in Table \ref{table:Bayesianoverview}. The model with the highest Bayesian evidence is highlighted in bold.}}
\end{table}

\noindent The calculation of the Bayesian evidence takes model complexity into account. A more complex model with a parameter space of higher dimensionality that converges to the same solution as a more basic model has a lower Bayesian evidence as a result. This inbuilt characteristic of multi-nested sampling retrieval becomes evident in the application of MERC to the mid-transit data. An inspection of the transmission spectrum shows a surprisingly deep sodium feature that is significantly broadened with no evident net blue or redshift and a symmetric line shape. 
\noindent Based on this assessment, we explored four likely scenarios for the line broadening. The most basic model is the isothermal case where the line broadening is solely attributed to temperature and pressure broadening with no atmospheric dynamics. In the three other cases, the isothermal case is combined with either two different types of zonal, horizontal winds or a vertically flowing, radially outwards bound wind. These three types of wind patterns are marked as $\mathrm{dtn}_{\cos\theta}$ for the day-to-night side flow, $\mathrm{srot}_{\cos\theta}$ for the super-rotational flow where the atmosphere rotates faster than the planet itself, and $\mathrm{ver}$ for the vertical, radial flow. $\cos\theta$ denotes the solid body type of zonal flow, where the strongest wind speed is located at the equator and no wind is encountered at the poles, mimicking a more physical wind pattern than a homogeneous distribution that creates unphysical artifacts at the poles. For more information on this implementation, see \citet{Seidel2021}.

For the mid-transit data, no evidence of either a day-to-night side wind, nor a super-rotational jet is found, with the basic model of no winds with a constant temperature profile as the preferred fit. Both the $\mathrm{dtn}_{\cos\theta}$ and $\mathrm{srot}_{\cos\theta}$ model converge to wind speeds around $0$ and the same temperature range as the basic model. This lead to the conclusion that the difference in evidence is based on the punishment of more complex models due to the increase in parameters and in fact, neither wind pattern has any impact on fit. The vertical wind pattern stands out with strong evidence compared to the other models. For an overview of the Bayesian evidence of each model and their strength of evidence based on the Jeffreys' scale, see Table \ref{table:comparison_mid_transit}. The posteriors and best-fits for all models can be found in Appendix \ref{app:postditcentr}. The vertical wind has its best-fit at a wind speed of $31.3\pm 6.7\, \kms$ with a retrieved isothermal temperature of $4093\pm229~K$ and the continuum parameter at $-2.99\pm 0.07$. The large uncertainty most likely stems from the reduced S/N in the dataset due to the split of the transit in a mid-transit and egress dataset. The strong wind speed, as well as the high temperature indicated that most likely higher layers of the atmosphere penetrating the lower boundary of the thermosphere are probed. Considering the wide range of altitudes probed, a temperature gradient cannot be ruled out and the isothermal profile is an approximation instead of the preferred temperature profile. The integration over the terminator, especially during mid-transit, automatically implies that we cannot distinguish between the day and night side contribution. The retrieved temperature of $\sim 4100\pm200~K$ indicates that indeed lower pressure regimes are probed where the temperature inversion is present \citep{MikalEvans2022}. Considering that the dominant wind pattern of radially outwards blowing wind is also attributed to higher atmospheric layers, the driver is higher atmospheric layers.

Additionally, as another intrinsic limitation of transmission spectroscopy from the ground, we loose all information on the probed pressure ranges and operate on relative pressure to an arbitrarily set base pressure only \citep[see ][ for more information]{Birkby2018,Seidel2020}. Considering from the wind pattern and the line depth, the assumption that the mid-transit data is able to probe down to our base pressure (the white light radius associated pressure level) is optimistic. It is more likely that the mid-transit dataset probes a lower pressure range higher in the atmosphere, especially considering that the mid-transit dataset probes large parts of the cooler night side where clouds might be present. As a consequence of this pressure uncertainty, the absolute value of the temperature can only be seen as an upper limit set by the line depth from the continuum. A possible avenue to increase the accuracy of the temperature estimation in the possible presence of clouds is the combination with low-resolution data to set the lower pressure boundary \citep[e.g. as in ][]{Pino2018} or to use estimations of the relative sodium abundance to break the pressure abundance degeneracy in multi-line studies \citep[e.g.][]{Pino2018, Allart2020, Maguire2022}.
Both approaches go beyond the scope of this paper and will be left for future work.

\subsection{Application to egress data}

\begin{table}
\caption{Comparison of the different models for the egress data.}
\label{table:comparison_egress}
\centering
\begin{tabular}{l c c l}
\hline
\hline
Model & $|\ln\mathcal{Z}|$   &  $|\ln\mathcal{B}_{01}|$ & Strength of evidence  \tablefootmark{a}  \\
\hline
$\mathrm{srot}_{\cos\theta}$  &  $1818.04\pm0.11$   & -  & - \\
$\mathrm{dtn}_{\cos\theta}$  &  $1818.10\pm0.11$   & $0.06$  & Inconclusive \\
isothermal  &  $1822.71\pm0.08$   & $3.66$  & Moderate evidence \\ 
ver  &  $1824.82\pm0.10$   & $6.78$  & Strong evidence\\
$\mathrm{\textbf{jet}}_{\cos\theta}, \mathrm{\textbf{ver}}$  &  $1825.88\pm0.10$   & $7.84$  & Strong evidence \\
\hline
\end{tabular}
\tablefoot{\tablefoottext{a}{ The base model to calculate $|\ln\mathcal{B}_{01}|$ is the isothermal model with no added wind patterns. The comparison stems from the Jeffreys' scale in Table \ref{table:Bayesianoverview}. The model with the highest Bayesian evidence is highlighted in bold.}}
\end{table}

\begin{figure*}[htb!]
\resizebox{\textwidth}{!}{\includegraphics[trim=2.0cm 0.0cm 2.0cm 0.0cm]{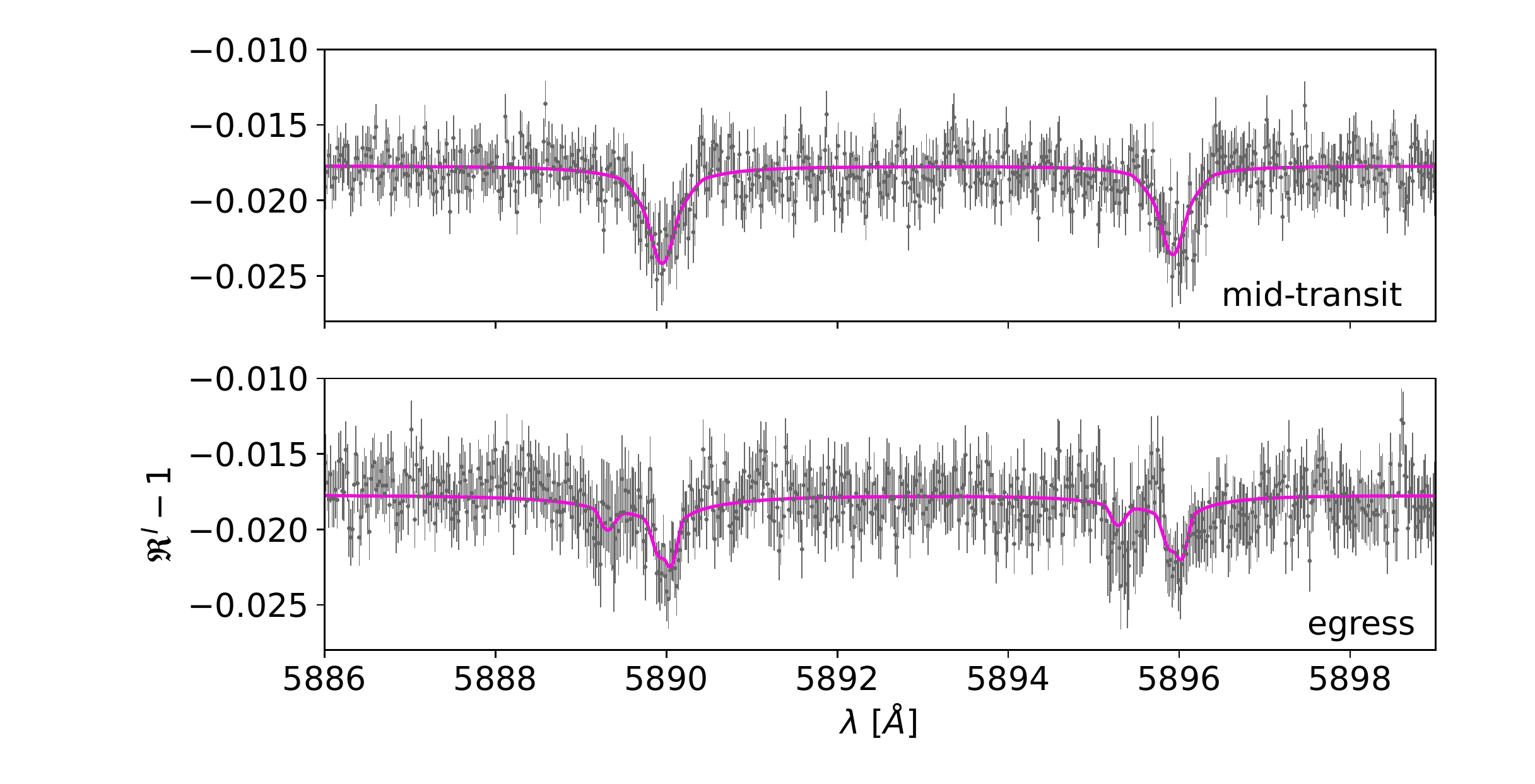}}
        \caption{Best-fit for both the mid-transit centred data and the egress data (top and bottom panel respectively). For the mid-transit data, the best-fit is a vertical, radial wind throughout the probed atmosphere, for the egress data the best-fit is achieved with an added jet in the lower atmosphere with $\cos\theta$ dependence and a vertical wind in the upper atmosphere. The best-fit models are shown in fuchsia on the dataset in gray.}
       \label{fig:dtn_verbestfit}
\end{figure*}

For the egress data, we chose the same models and priors as for the mid-transit data and added one model with an equatorial jet over the evening terminator as described in Section \ref{sec:jet}. The Bayesian evidences as derived with MERC are shown in Table \ref{table:comparison_egress}, the posteriors for each model together with the best-fit can be found in Appendix \ref{app:postditegress}. The egress data, together with the best-fit model, is shown in the bottom panel of Figure \ref{fig:dtn_verbestfit}. Just as for the mid-transit centred data, both a super-rotational wind or a day-to-night side wind over both terminators and at all altitudes do not provide a good fit to the data. Due to their higher complexity the isothermal basic model is preferred with moderate evidence. The main peak of the sodium doublet remains broadened compared to the instrument's profile, which is reflected in the strong evidence for the vertical, radial wind model and the best-fit for both the vertical, radial wind model at high velocities instead of close to zero for both the case of only radial winds and the model including a jet-like feature.

\noindent To account for the high-velocity absorption components on the blueshifted side, we created a two-layer model with a one-sided jet in the lower atmosphere and vertical, radial wind for the central broadening in the upper atmosphere. We base this strategy on two arguments: from global circulation models (GCMs) we know that jets are usually restricted to bands along the equator \citep{Showman2009,Showman2018,Parmentier2018}. From previous work on WASP-76~b, an ultra-hot Jupiter with many of the same characteristics as WASP-121~b especially in terms of temperature \citep{West2016}, we know that its lower atmosphere is dominated by day-to-night side winds across both terminators. However, as we do not have resolved data in latitude, a day-to-night side wind over a broad latitude range is degenerate with a multi-jet scenario as seen in Jupiter's atmosphere \citep{LiuSchneider2010}. This two layer approach requires a switch between wind patterns at a specific point in the vertical structure of the atmosphere. In \cite{Seidel2020}, it was shown that introducing an additional parameter for this switch is not conducive to a better convergence. In MERC, the atmospheric structure in altitude is implemented via a log-scale decrease in pressure from a pre-defined 'surface' where the planetary atmosphere becomes opaque. For consistency with existing literature, this surface pressure at the continuum level of the transmission spectrum is set at $10$ bar \citep[e.g.][]{LecavelierdesEtangs2008, Agundez2014,Line2014, Pino2018}. For the WASP-121~b equivalent, WASP-76~b, \citet{Seidel2021} explored various different altitudes for the switch in wind patterns ($10^{-3}$ bar, $10^{-4}$ bar, and $10^{-5}$ bar). The Bayesian evidence indicated that a switch at $10^{-3}$ bar is preferred with strong evidence independently of what zonal winds are applied in the lower atmosphere. This pressure is also the upper boundary explored by Global Circulation Models for zonal winds \citep{Showman2009}. We have as a consequence set the change from zonal winds to radial winds at the same pressure level for the two layer model applied here to the egress data of WASP-121~b.

\begin{figure}[htb!]
\resizebox{\columnwidth}{!}{\includegraphics[trim=1.0cm 0.0cm 1.0cm 0.0cm]{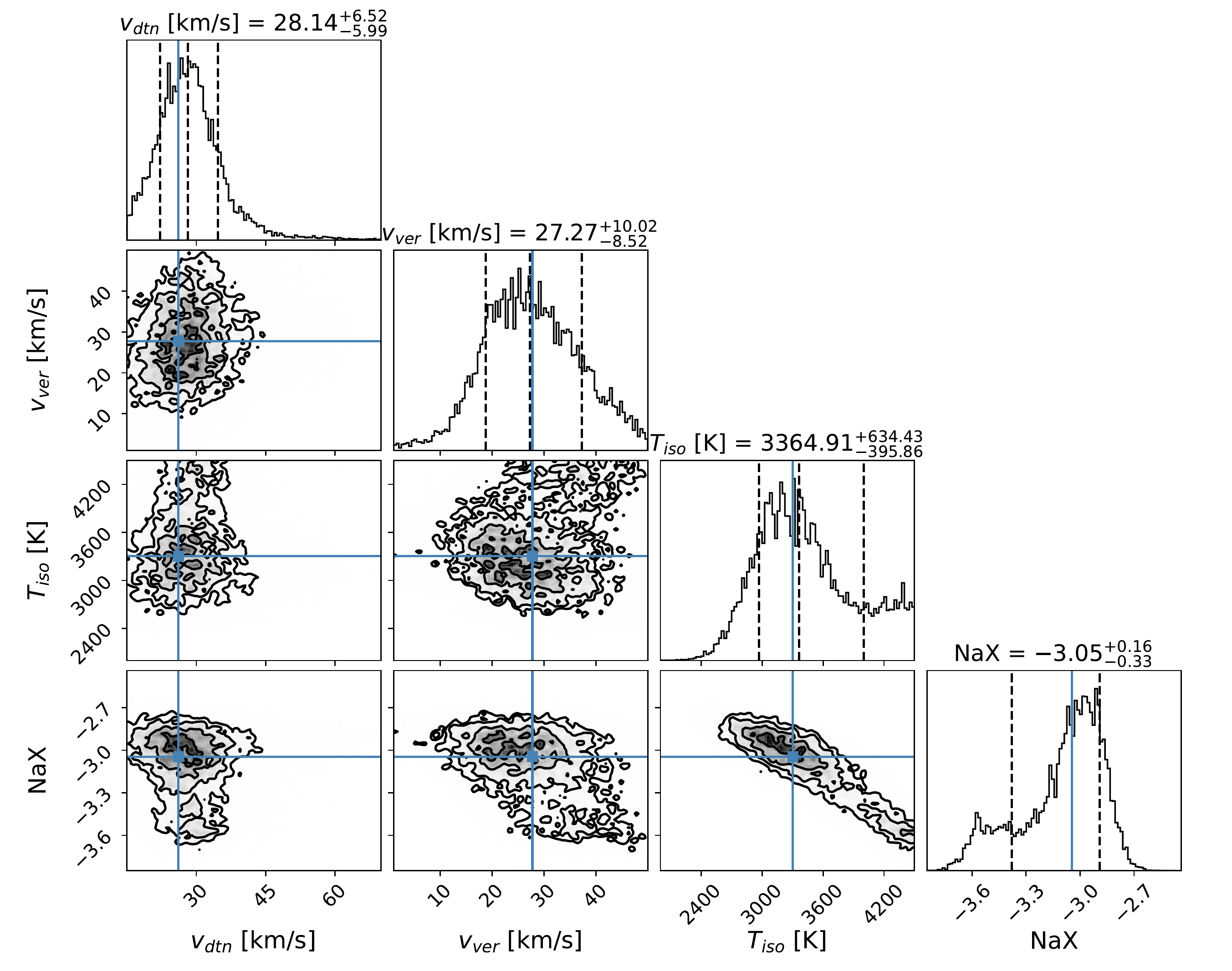}}
        \caption{Posterior distribution of isothermal line retrieval with an added jet in the lower atmosphere across the evening terminator and a vertical wind in the upper atmosphere. The change of layers was set at $p=10^{-3}$ bar, with the pressure at the surface set to $P_0=10$ bar.}
        \label{fig:dtn_verbestposterior}
\end{figure}

\noindent The model including the one-sided jet is preferred over all other models. The next closest model on the Jeffreys' scale is the vertical wind without jet with a difference of Bayesian evidence of $|\ln\mathcal{B}_{01}|=1.06$ which corresponds to a better fit probability from adding the jet of over $75~\%$ (see Table \ref{table:Bayesianoverview} for more information on the Jeffreys' scale). MERC compares the data over the entire wavelength range of the sodium doublet, which means that in comparison the high-velocity feature only impacts a rather small part of the overall data than the much larger broadened main sodium doublet. It is, therefore, not surprising that the difference between the preferred model with the jet and the model with only the vertical wind is not larger given the dominance of the broadened sodium doublet. The posterior distribution with the best-fit parameters marked in blue for the preferred model (jet and vertical wind) is shown in Figure \ref{fig:dtn_verbestposterior}. 
The jet velocity is retrieved at $26.7~\kms$ with a one sigma uncertainty of $\pm 7.2~\kms$, the vertical wind velocity in the upper atmosphere at $27.8~\kms$ with a one sigma uncertainty of $\pm 9.3~\kms$. The best-fit temperature is $3350\pm470~K$ and the continuum parameter is retrieved at $-3.1\pm0.2$.
The retrieved temperature is compatible with the results from \citet{MikalEvans2022} where the day-side shows lower atmospheric temperatures of up to $4000~K$ above $10^{-3}$~bar. The results from \citet{Maguire2022} with a retrieved temperature for the day-side of $3450\pm160$~K also stem from high-resolution ESPRESSO data and show that both retrieval approaches produce the same T-P profile with a remarkable consistency in the retrieved temperature.

\section{Discussion and Conclusions}
\label{sec:diss}

We reanalysed the partial ESPRESSO 4-UT mode transit observations of the ultra-hot Jupiter WASP-121~b first presented in \citet{Borsa2021} to study the resolved sodium doublet. We show that a prominent blueshifted high-velocity absorption component seen for both doublet lines stems only from the egress half of the in-transit data by separating the transit in two distinct datasets, the mid-transit centred and the egress data. We assess the two most important impacts on the resolved line shape, Doppler smearing caused by the planetary movement during each exposure and the RM-effect. Following \citet{Borsa2021}, we show that the RM- and CLV-effect influences the line shape, but that it cannot account for the blueshifted feature. The effect of the Doppler-smearing is estimated at approximately one resolution element and, therefore, has no impact on the  line shape beyond the instrumental limits.

We apply MERC to both datasets, mid-transit centred and the egress, and show that the broadening of the mid-transit sodium doublet is best retrieved by a vertical, radial wind in the upper atmosphere at $34\pm7~ \kms{}$. The large uncertainty is due to the reduced S/N from splitting the partial transit into two parts as well as the accounted Doppler-smearing from Section \ref{sec:dataset}. For the egress data, the best-fit is achieved with a vertical wind at $27.8\pm9.3~\kms{}$ above $10^{-3}$ bar and a jet below that threshold extending across $60^\circ$ in latitude on the evening limb with a wind velocity of $26.8\pm7.3~\kms{}$. The wind model including the jet is preferred over the model with only the vertical wind albeit with a small difference in Bayesian evidence.
The localised but strong jet along the equator displaces enough sodium from the expected location in the transmission spectrum via Dopplershifting to account for the blueshifted high-velocity absorption component we have first analysed visually on the resolved lines in \citet{Borsa2021}. Unfortunately, the 4-UT data lacks the ingress part of the data which means it is unclear whether this jet feature only exists on the evening terminator or if it is part of a global day-to-night side wind.

\subsection{Constraints on the location of the high-velocity jet}

Under the caveat of the difference in Bayesian evidence, in the following we will explore the possible location and origin of the jet stream. We plan to observe the ingress of this particular system to further establish the observational evidence and for modelling efforts to include the jet in GCM simulations.
The blueshifted high-velocity absorption component is seen for both sodium doublet lines and, at similar blueshift from line centre, also for the $H-\alpha$ line \citep[see Figure 9 in][]{Borsa2021}. For the resolved K and $H-\beta$ lines a similar feature can not be confirmed due to the noise level \citep[see Figure 11 and 12 in][]{Borsa2021}. There is no hint of a blueshifted high-velocity absorption component for Li \citep[see Figure 11 in][]{Borsa2021}, however, all of the mentioned resolved singular lines probing below the altitudes traced by the sodium doublet (Li, K, and $H-\beta$) show a pronounced blueshift from the expected line centre. While a rigorous analysis of the blueshift and the similar feature in the H-$\alpha$ line is beyond the scope of this paper and the current capabilities of MERC, the visual inspection allows us to estimate the altitude at which this feature is generated under the strong caveat of a homogeneous continuum level for all these lines. Due to the loss of the planetary continuum \citep{Birkby2018} this approach can only be seen as a first order estimate. The 4-UT sodium feature probes up to approximately 1.15 relative planetary radii ($R_p$), while Li probes only to 1.08 $R_p$ \citep[see Figure 11 and Table 2 in][]{Borsa2021}. As a consequence, the blueshifted feature is most likely generated at an altitude in the range of [1.08,1.15] $R_p$. Under the assumption of a constant base pressure and a hydrostatic atmosphere this altitude corresponds to the higher layers of the atmosphere above $10^{-2}$~bar, which corresponds to the start of the temperature inversion as observed in previous work. The first order estimate of the origin of the jet is thus compatible with the retrieved temperature regime and we probe the atmosphere in its heated temperature inversion at the base of the thermosphere.

Using the cross-correlation technique on the iron lines to probe deeper layers of the atmosphere, \citet{Borsa2021} estimates that the Fe I lines probe the atmosphere up to 1.05 $R_p$ just below the range we estimate for the origin of the blueshifted high-velocity absorption components. Despite the large uncertainty of the retrieval results of the wind speeds for the jet, the wind speeds probed by the Fe I tracing are below the jet wind speed by a factor of 2-3. A possible explanation for this discrepancy is that different, but connected atmospheric layers are probed. The trend for next to no day-to-night side wind movement for the mid-transit data but a comparatively strong wind in the second half of the data observed for Fe I in \citet{Borsa2021} is also observed for sodium with MERC. However, Fe I only probes up to 1.05 $R_p$, just below our estimate for the origin of the jet feature at [1.08,1.15] $R_p$. This could mean that Fe I probes the lower layers of the jet deeper in the atmosphere, whereas sodium is able to probe the central core of the jet at lower atmospheric density where stronger wind speeds are possible.

\begin{figure}[htb]
\resizebox{\columnwidth}{!}{\includegraphics[trim=0.0cm 1.0cm 0.0cm 1.0cm]{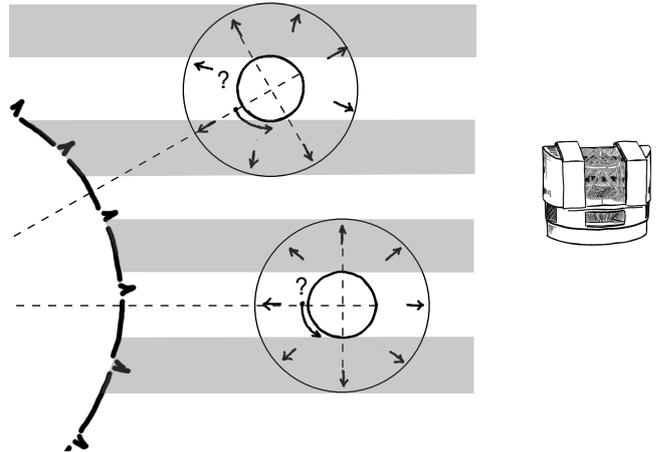}} 
	\caption{Graphic of atmospheric parts probed during the mid-transit and egress data in a polar planetary view from Figure \ref{fig:sketch_polar} updated with the wind patterns. In this polar view, the radial, vertical wind is seen in the same way during all in-transit exposures, but the high-velocity jet in the lower atmosphere is only visible during egress. This means its extension cannot reach the evening limb during mid-transit, as it would otherwise also influence the mid-transit data. Given that we have no ingress data, we do not know if the jet is also moving from the day to the night side over the other terminator, is part of a super-rotation, or is not present at all. This missing part of information is indicated with a question mark.}
	\label{fig:sketch_windPatterns}
\end{figure}

Another interesting aspect is the change in visible atmosphere from the mid-transit to the egress data, which is sketched for both the mid-transit and the egress part of the data in Figure \ref{fig:sketch_windPatterns}. In Section \ref{sec:dataset}, we estimate from the orbital phases that the opening angle for the egress data varies from $10^\circ$ to $30^\circ$ with a mean opening angle of $20^\circ$. Given that no indication of the jet is visible for the mid-transit dataset stretching from $-10^\circ$ to $10^\circ$, the jet is most likely starting to be visible across the terminator between an opening angle of $10^\circ$ to $20^\circ$ to still generate enough S/N in the integrated transmission spectrum. Both \citet{MikalEvans2022} and \citet{Daylan2021} did not observe a significant difference between the brightest and the substellar point from HST observations, as well as in \citet{Bourrier2020} from TESS observations, indicating that the hotspot, and thus the most likely source of a strong jet stream, is set at the substellar point and not offset. Additionally, they found a strong temperature difference between the day and night side, which they attribute to insufficient heat transportation and two distinct realms on the day and night side. Taking into account these results, it is likely that the jet extends from the substellar point at $90^\circ$ from the evening terminator to [20,40]$^\circ$ after which it is abruptly dispersed, which is consistent with the HST observations in \citet{Evans2017}. This leads us to the conclusion that our observations for the sodium doublet in WASP-121~b are compatible with a supersonic, equatorial jet that is dispersed before it reaches the night side of the atmosphere. 

\subsection{Possible origins of the observed jet and caveats}

A similar observation of a high-velocity feature was made from the CCF centroids in \citet{Cauley2020} for the ultra-hot Jupiter WASP-33~b, however in their analysis they rule out a supersonic, localised jet as un-physical despite the good fit to their observations. 
Considering that we have constrained the origin of this jet to the upper atmosphere between [1.08,1.15] $R_p$, a supersonic jet is far from un-physical. Supersonic jets in planetary atmospheres can be produced by the excitation of a stationary planetary scale Rossby wave by stellar irradiation \citep{ShowmanPolvani2010, ShowmanPolvani2011} as is likely for tidally-locked planets such as WASP-121~b \citep[see also ][ for a treatment of 3D effects]{Tsai2014}. While these supersonic jets can remain stable across the atmosphere for higher pressures, in the lower pressure regime that is applicable to the here observed jet (P < 10 mbar), supersonic jets are susceptible to large scale barotropic Kelvin- Helmholtz instabilities and short scale vertical shear instabilities \citep{Fromang2016}. This scenario is also supported by the results on the temperature structure of the atmosphere.

In this context it is important to mention a degeneracy in the interpretation of the high-velocity feature as a jet. Due to the opening angle we probe part of the hotter dayside with the evening limb where we see the origin of the high-velocity feature. However, additionally, we probe more of the night side on the morning limb. A day-to-night side wind rotates with the planetary rotation on the evening limb, but provides a Doppler shift against the planetary rotation on the morning limb. In theory, the high-velocity feature could be produced on the night side of the morning limb only or result from a combination of both jets at exactly the correct wind speed difference to offset the planetary rotation. We do not explore these two scenarios further due to the following reasoning:
Assuming that the high-velocity feature is produced by two arms of the jet in both the parts probed on the day-side on the evening limb and the night-side on the morning limb, the sharpness of the feature with little smearing would mean a near perfect equilibrium between the two components, always offsetting the difference induced by the planetary rotation. The planetary rotation of WASP-121~b is approximately $7.4~\kms{}$, resulting in a needed offset of $\sim 15~\kms{}$ for all different contributions over the two limbs at all times and longitudes. While we cannot rule out this coincidental scenario with the current data, it seems unlikely that the night-side jet has velocities exactly $\sim 15~\kms{}$ larger than the day-side. Additionally, if the jet only stems from the probed part of the night side, its velocity would translate to a needed jet velocity of over $\sim 41\pm7~\kms{}$ to offset the different line of sight velocity of the planetary rotation, significantly larger and harder to produce and explain at the probed altitude levels. Additionally, the night side should have a temperature below $2000~K$ \citep{MikalEvans2022,Bourrier2020} while the presented work probes atmospheric layers with temperatures exceeding $3000~K$. 

 For similar results of a day-to-night side wind for WASP-76~b in \citep{Ehrenreich2020, Kesseli2021, Seidel2021}, the important caveat was raised that a temperature gradient and strong heat transportation across the terminator could create a similar offset \citep{Wardenier2021}. Low-resolution observations over different phases have ruled out this scenario \citep{Evans2017}, showing insufficient heat transportation across the terminator and two distinct regimes on the day and night side. Our hypothesis of a disrupted jet due to instabilities is based on observations of the sodium doublet. It is in theory possible that we simply do not trace the super-sonic jet further due to the condensation of sodium. Under the assumption that the jet is visible on the hotter day-side, additional ionisation of sodium when crossing to the night side can be ruled out. The condensation scenario, which was proposed for iron in WASP-76~b \citep{Ehrenreich2020}, would require temperatures at approximately $1000~K$ for half of the atmospheric sodium to condensate to the liquid phase \citep{Wood2019}. Considering night side temperatures estimated as the lower boundary at $1500~K$ in the lower atmosphere \citep{MikalEvans2022}, it is unlikely that a large fraction of sodium is condensated. While we cannot rule out that our observations trace the temperature gradient and condensation instead of a true disruption due to atmospheric instabilities, it is an unlikely scenario. 
 
Given the high altitudes probed, magnetic drag becomes more important \citep{Beltz2022}. Recent work including magneto-hydrodynamics (MHD) in GCMs has shown that magnetic drag can disrupt the often claimed super-rotational jets of ultra hot Jupiter atmospheres \citep{Beltz2022}. Given the observational nature of this paper including MHD is beyond its scope. The application of global circulation models including MHD within the observational constraints provided here would allow explore the temperature gradient vs Kelvin-Helmholtz instability explanation from first principles and we encourage such endeavours by other groups.

 \subsection{Future avenues}

Unfortunately, the 4-UT transit is only partial and we lack the information at ingress to search for the other stream of a possible day-to-night side jet over both terminators. We can, therefore, not say whether the observations show a supersonic, super-rotational jet or one stream of a supersonic day-to-night side wind. Additionally, we assume that the planet rotates perpendicular to its orbital plane with the maximum rotational velocity in the line of sight. This is a reasonable assumption given the tidally-locked nature of ultra-hot Jupiters, but will become a major caveat when this technique is applied to smaller exoplanets further out from their host stars.

An important recent avenue to observe atmospheric dynamics in ultra-hot Jupiters is the study of emission lines at various phases which provides information on the parts of the day-side not accessible during transit \citep[see ][ for the first detection of iron in emission]{Pino2020}. These emission lines can then be mapped to a likelihood function using the CCF technique and thus give us insights into the line depths and Doppler-shifts throughout different phases as first explored in \citet{Pino2022}. The time-resolved information of the day-side, especially when taking magnetic draft into account \citep[see e.g.][]{Beltz2022b}, provides the information missing from the day-side not accessible during transit to create a more holistic observational picture of exoplanet atmospheric circulation.

As a future avenue for in-transit observations, we emphasise the need for 4-UT data and several 4-UT transit observations of the same target for a finer phase resolution in resolved line studies. Due to the inherent three-dimensional nature of atmospheres, atmospheric dynamics and thermal or chemical gradients across the terminator can strongly influence the spectral shape. This behaviour can only be recovered by time-resolving transits and retrieving the evolution of the spectral shape in time. As shown recently for CCF tracing in \citet{Gandhi2022}, a time-resolved approach is far superior in fitting the data than 1D time-integrated studies.

The combination of various resolved lines, resolving the issue of varying continuum levels, would allow us to more robustly trace the changes in atmospheric dynamics in transit over a wider range of altitudes than only the line shape and wings of the sodium doublet \citep[see e.g.][as the current state of the art]{Zhang2022}. Additionally, the inclusion of various lines will allow the parallel retrieval of relative abundances \citep[see e.g.][]{Gibson2022,Maguire2022} which in turn informs a more accurate temperature pressure profile. The study of resolved line shapes will further mature as a tool to study the imprint of atmospheric dynamics as the data quality increases and will unfold its full potential in the era of Extremely Large Telescopes in the next decade.

%----------------------------------------------------------------------------------------
%       ACKNOWLEDGEMENTS
%----------------------------------------------------------------------------------------
\begin{acknowledgements}
The authors acknowledge the ESPRESSO project team for its effort and dedication in building the ESPRESSO instrument. The ESPRESSO Instrument Project was partially funded through SNSF’s FLARE Programme for large infrastructures. This work relied on observations collected at the European Organisation for Astronomical Research in the Southern Hemisphere. This work has been carried out within the framework of the National Centre of Competence in Research PlanetS supported by the Swiss National Science Foundation. The authors acknowledge the financial support of the SNSF. FPE and CLO would like to acknowledge the Swiss National Science Foundation (SNSF) for supporting research with ESPRESSO through the SNSF grants nr. 140649, 152721, 166227 and 184618. YA acknowledges the support of the Swiss National Fund under grant $200020\_172746$. JIGH, RR, CAP and ASM acknowledge financial support from the Spanish Ministry of Science and Innovation (MICINN) project PID2020-117493GB-I00. ASM, JIGH and RR also acknowledge financial support from the Government of the Canary Islands project ProID2020010129. CJM acknowledges FCT and POCH/FSE (EC) support through Investigador FCT Contract 2021.01214.CEECIND/CP1658/CT0001. R. A. is a Trottier Postdoctoral Fellow and acknowledges support from the Trottier Family Foundation. This work was supported in part through a grant from FRQNT. This work was financed by FCT - Fundação para a Ciência e a Tecnologia  under projects UIDB/04434/2020 \& UIDP/04434/2020, CERN/FIS-PAR/0037/2019 and PTDC/FIS-AST/0054/2021. The research leading to these results has received funding from the European Research Council (ERC) through the grant agreement 101052347 (FIERCE) and grant agreement 947634 (Spice Dune). This work was supported by FCT - Fundação para a Ciência e a Tecnologia through national funds and by FEDER through COMPETE2020 - Programa Operacional Competitividade e Internacionalização by these grants: UIDB/04434/2020; UIDP/04434/2020. J.L-B. acknowledges financial support received from "la Caixa" Foundation (ID 100010434) and from the European Unions Horizon 2020 research and innovation programme under the Marie Slodowska-Curie grant agreement No 847648, with fellowship code LCF/BQ/PI20/11760023. This research has also been partly funded by the Spanish State Research Agency (AEI) Project No.PID2019-107061GB-C61. We thank the anonymous referee for their kind comments, which improved the clarity of the main points of the manuscript.

\end{acknowledgements}

%----------------------------------------------------------------------------------------
%       REFERENCE LIST
%----------------------------------------------------------------------------------------
%
\bibliographystyle{aa} % style aa.bst
\bibliography{CulpaAlViento}
%----------------------------------------------------------------------------------------
%       APPENDIX
%----------------------------------------------------------------------------------------
%
\appendix
%\begin{appendix}
%\onecolumn
\section{Posterior distributions of the retrievals on the mid-transit data}
\label{app:postditcentr}

\begin{figure}[hbt]
\resizebox{\columnwidth}{!}{\includegraphics[trim=-0.0cm 0.0cm -0.0cm 0.0cm]{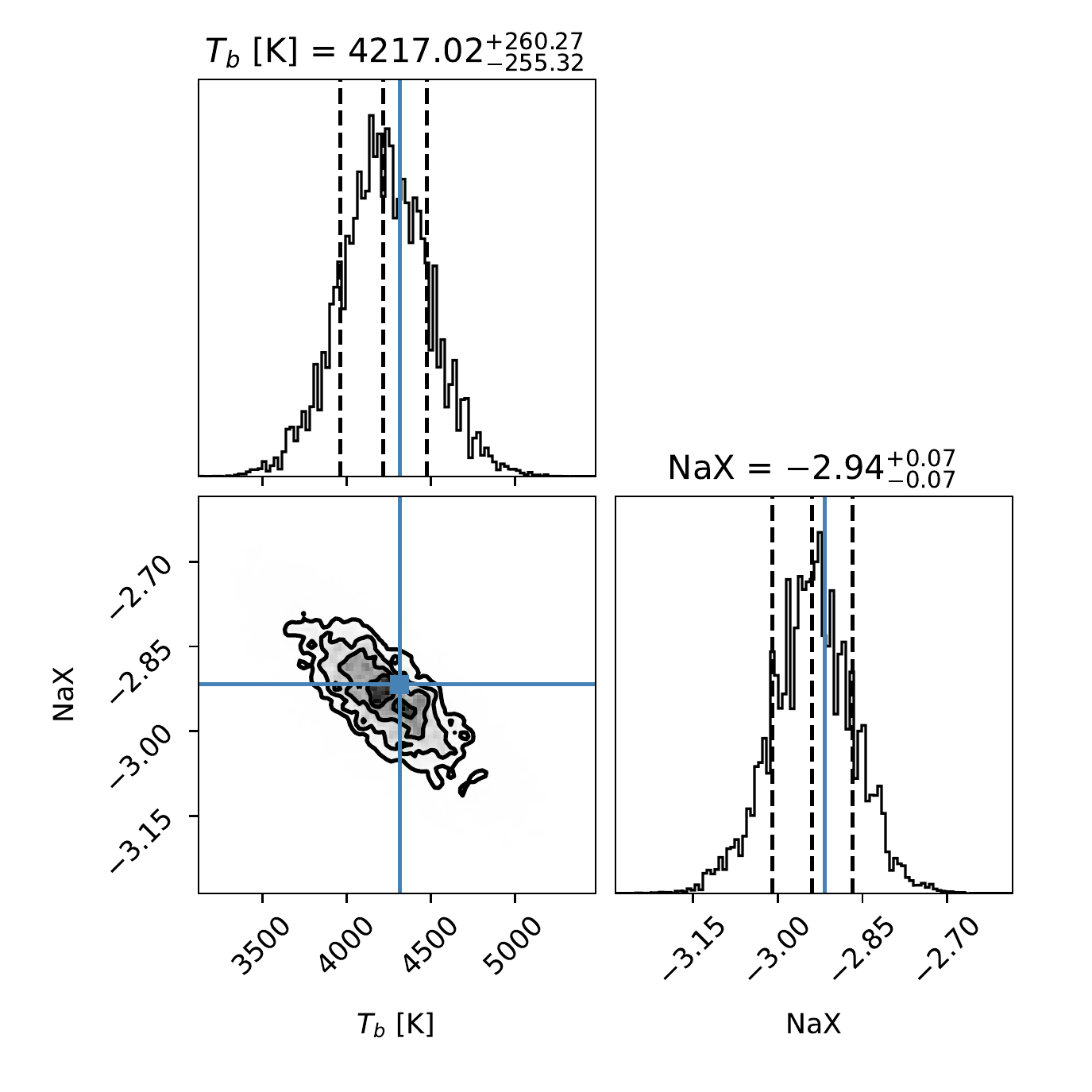}}
        \caption{Posterior distribution of isothermal line retrieval.}
        \label{fig:isoposteriorcen}
\end{figure}

\begin{figure}[htb!]
\resizebox{\columnwidth}{!}{\includegraphics[trim=0.0cm 0.0cm 0.0cm 0.0cm]{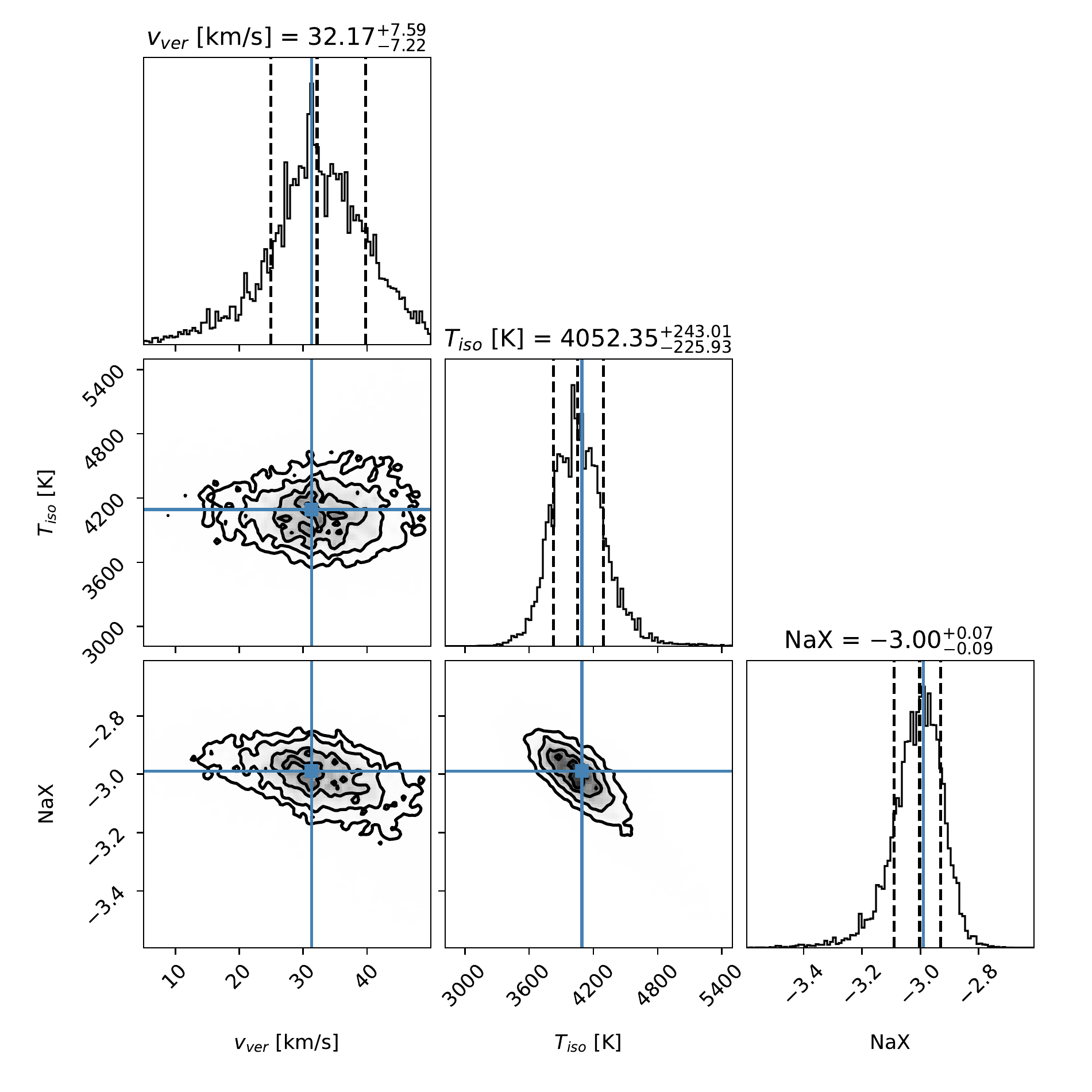}}
        \caption{Posterior distribution of isothermal line retrieval with an added vertical wind constant throughout the atmosphere.}
        \label{fig:vverticalposteriorcen}
\end{figure}

\begin{figure}[htb!]
\resizebox{\columnwidth}{!}{\includegraphics[trim=0.0cm 0.0cm 0.0cm 0.0cm]{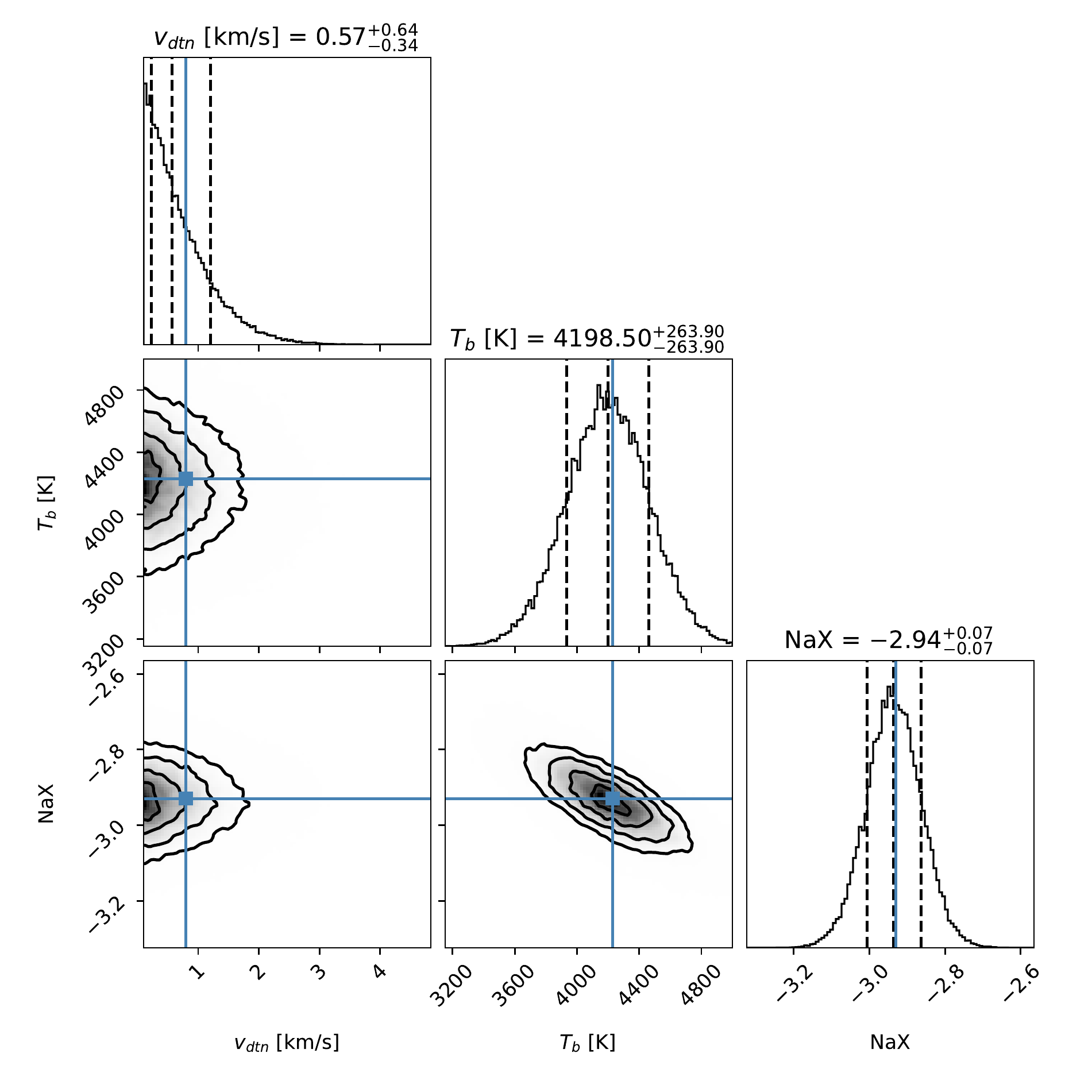}}
        \caption{Posterior distribution of isothermal line retrieval with an added day-to-night side wind constant throughout the atmosphere. }
       \label{fig:dtnposteriorcen}
\end{figure}

\begin{figure}[htb!]
\resizebox{\columnwidth}{!}{\includegraphics[trim=0.0cm 0.0cm 0.0cm 0.0cm]{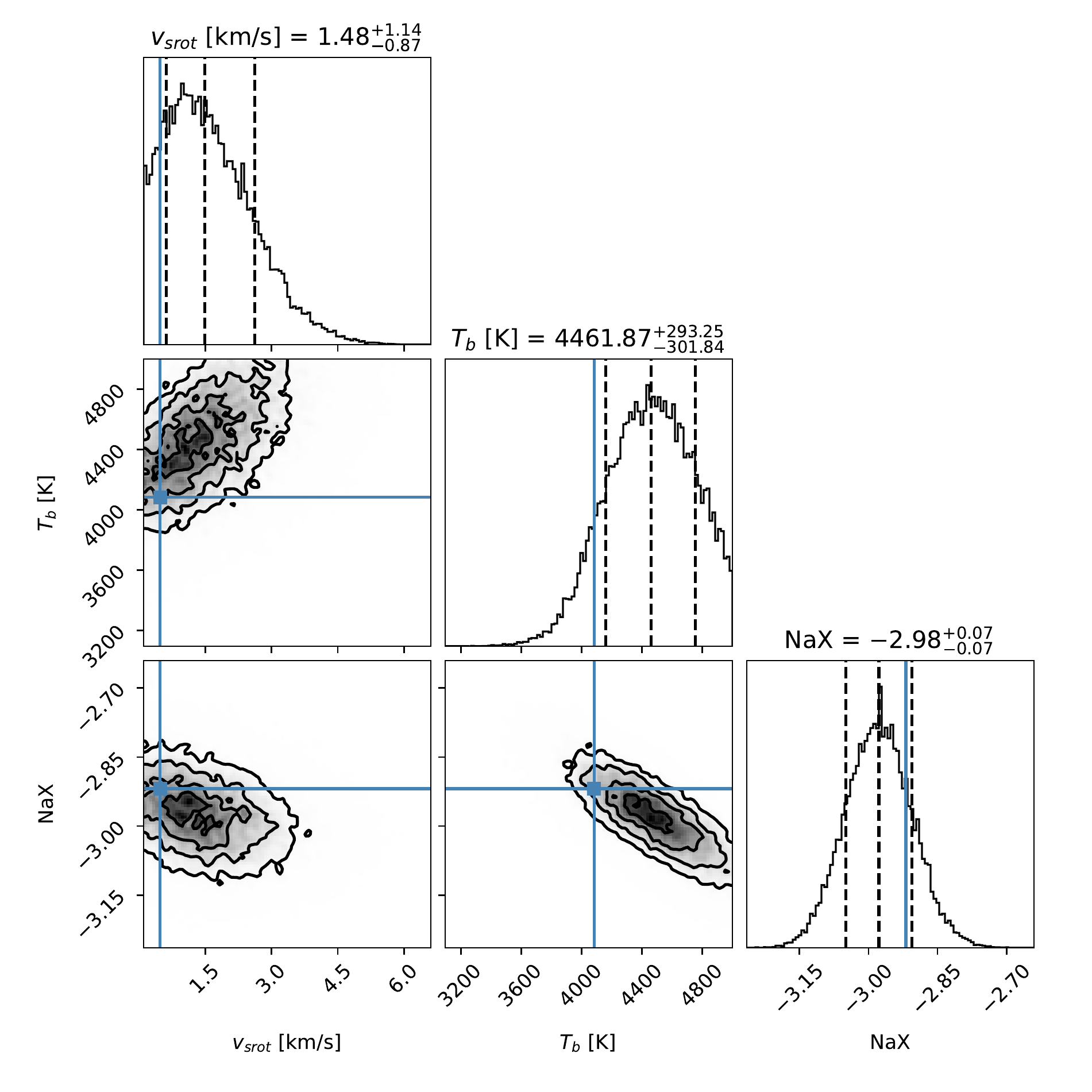}}
        \caption{Posterior distribution of isothermal line retrieval with an added super-rotational wind constant throughout the atmosphere.}
        \label{fig:superrotposteriorcen}
\end{figure}

\newpage
\section{Posterior distributions of the retrievals on the egress data}
\label{app:postditegress}

\begin{figure}[hbt]
\resizebox{\columnwidth}{!}{\includegraphics[trim=-0.0cm 0.0cm -0.0cm 0.0cm]{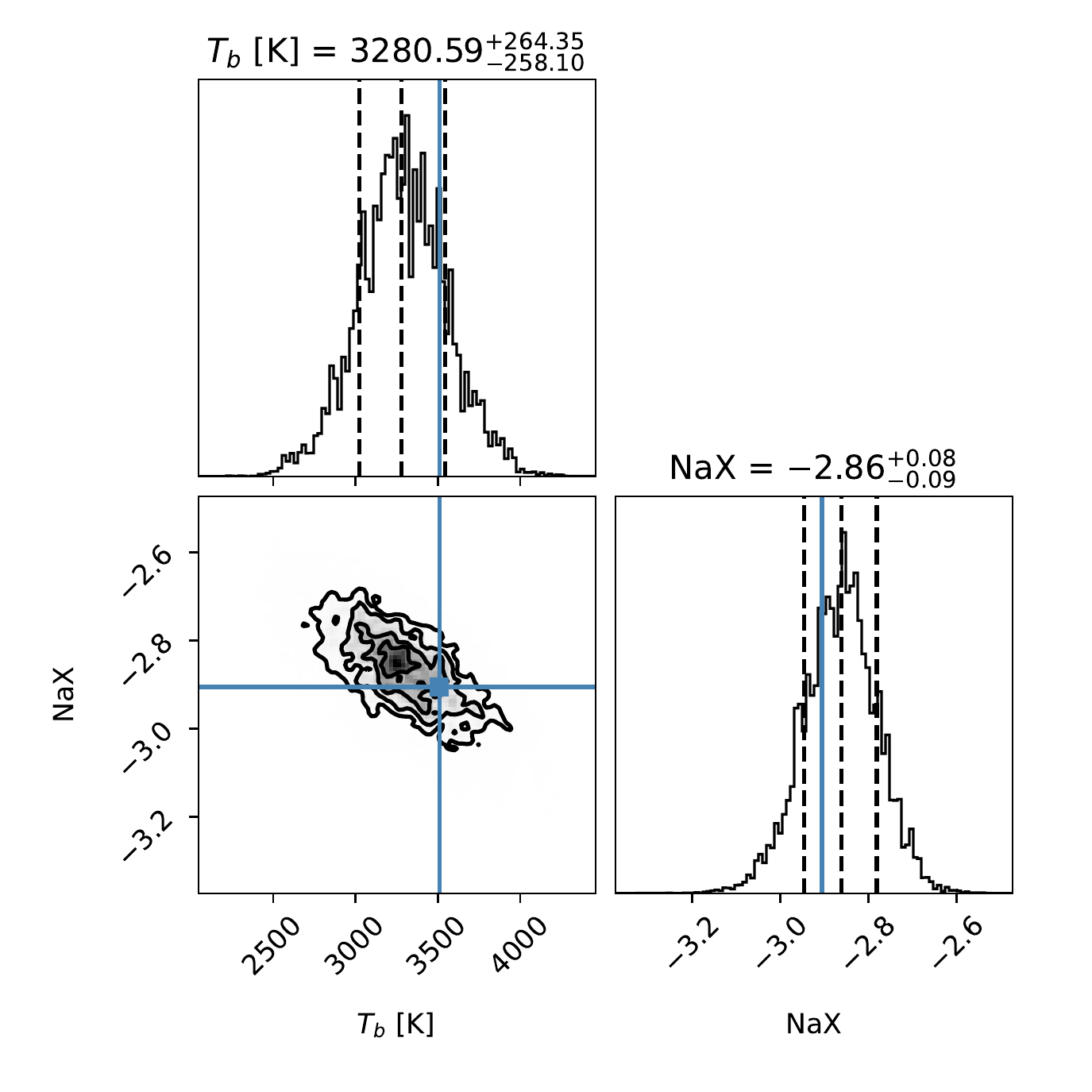}}
        \caption{Posterior distribution of isothermal line retrieval.}
        \label{fig:isoposterior}
\end{figure}

\begin{figure}[htb!]
\resizebox{\columnwidth}{!}{\includegraphics[trim=0.0cm 0.0cm 0.0cm 0.0cm]{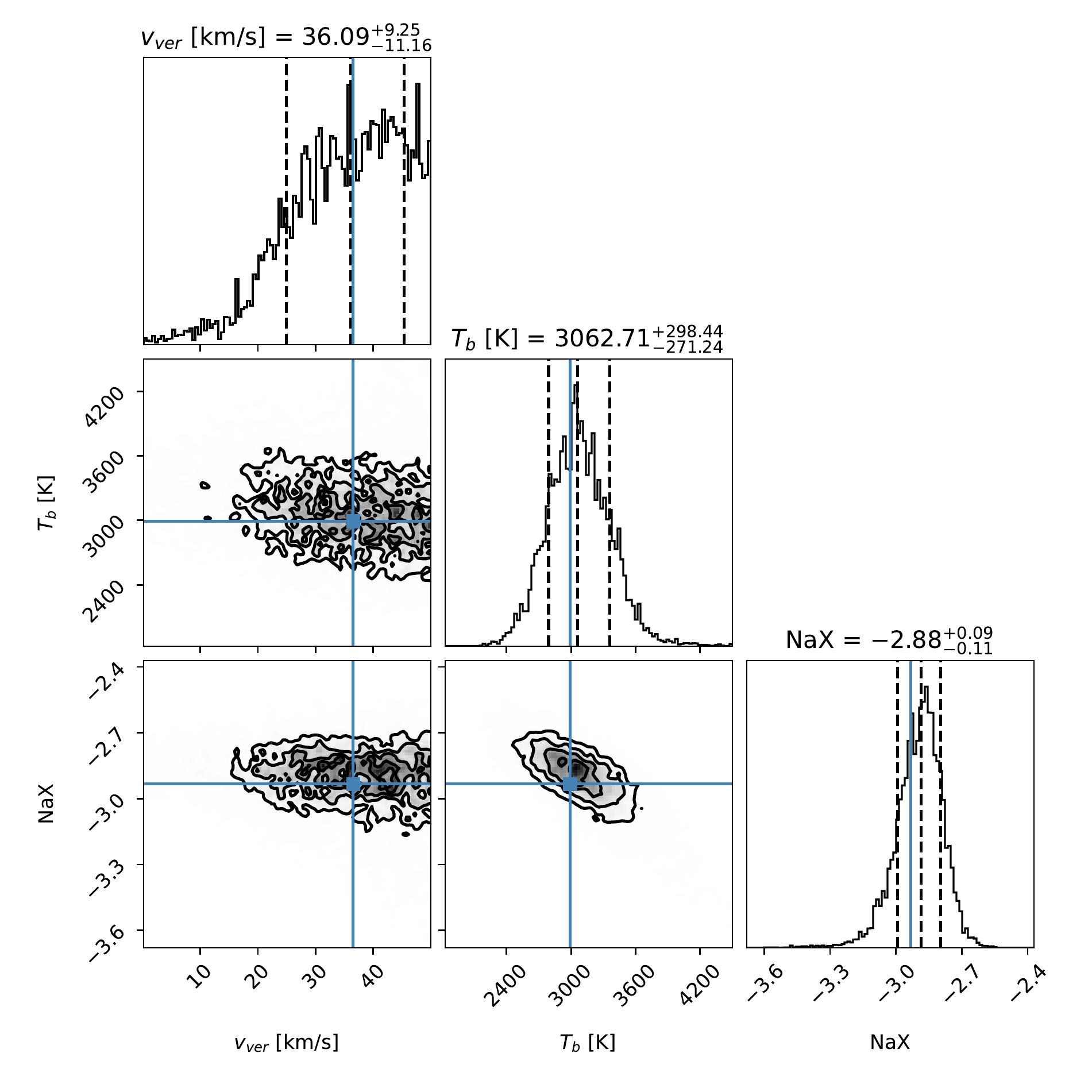}}
       \caption{Posterior distribution of isothermal line retrieval with an added vertical wind constant throughout the atmosphere.}
        \label{fig:vverticalposterior}
\end{figure}

\begin{figure}[htb!]
\resizebox{\columnwidth}{!}{\includegraphics[trim=0.0cm 0.0cm 0.0cm 0.0cm]{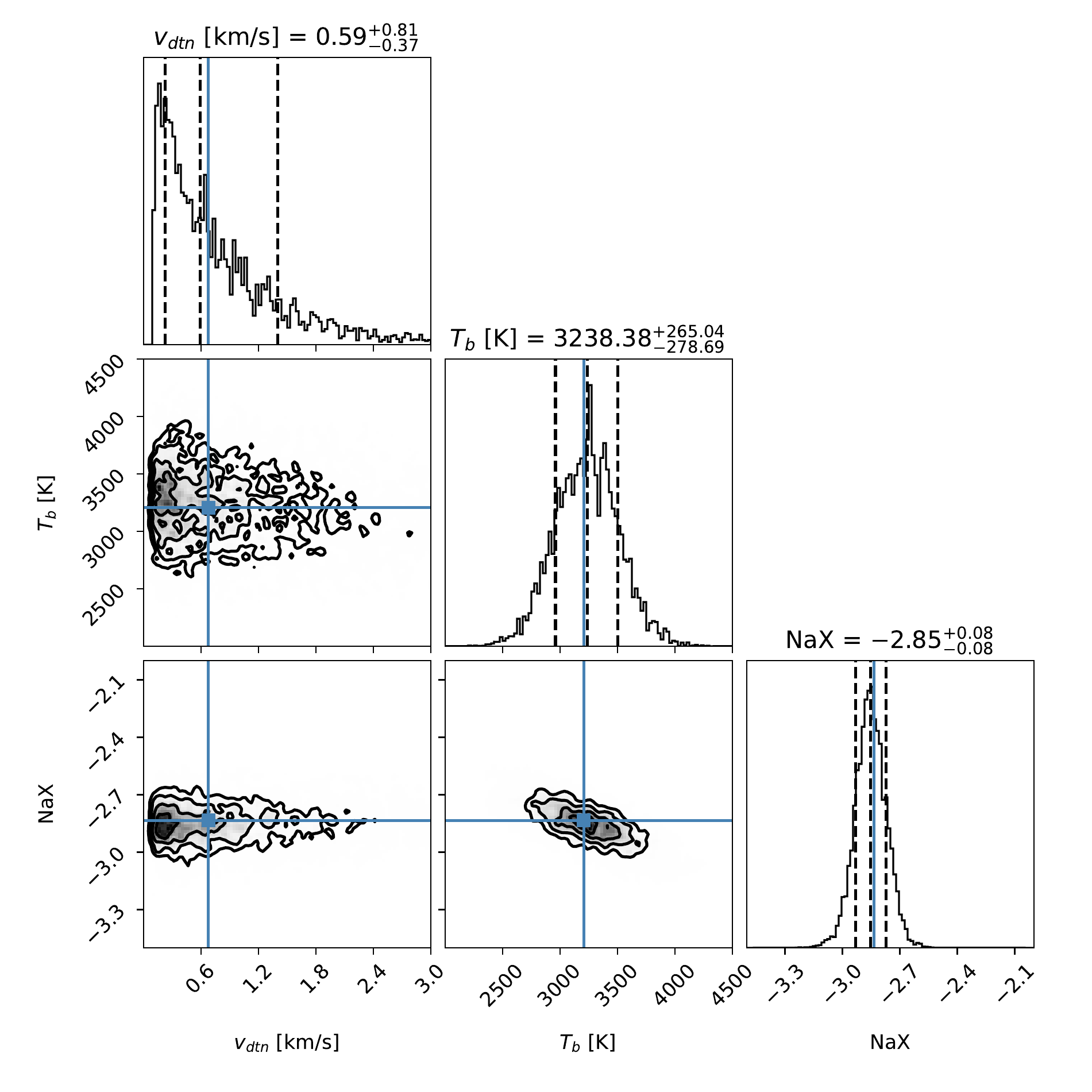}}
        \caption{Posterior distribution of isothermal line retrieval with an added day-to-night side wind constant throughout the atmosphere. }
       \label{fig:dtnposterior}
\end{figure}

\begin{figure}[htb!]
\resizebox{\columnwidth}{!}{\includegraphics[trim=0.0cm 0.0cm 0.0cm 0.0cm]{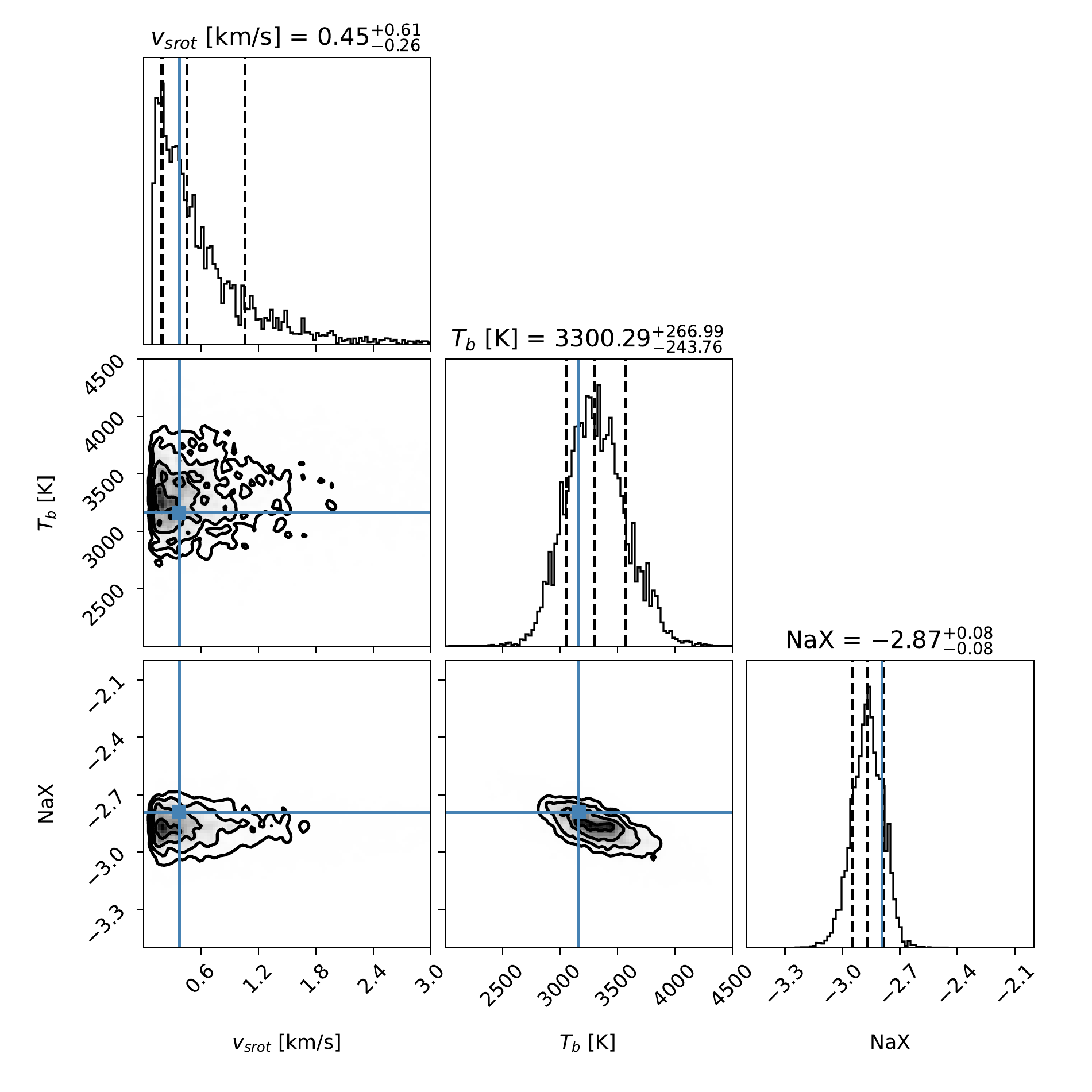}}
        \caption{Posterior distribution of isothermal line retrieval with an added super-rotational wind constant throughout the atmosphere. }
        \label{fig:superrotposterior}
\end{figure}

\end{document}